\documentclass[pra,showpacs,twocolumn,superscriptaddress]{revtex4}

\def\temp{1.35}%
\let\tempp=\relax
\expandafter\ifx\csname psboxversion\endcsname\relax
\else
    \ifdim\temp cm>\psboxversion cm
    \else
      \let\temp=\psboxversion
      \let\tempp= 
    \fi
\fi
\tempp
\let\psboxversion=\temp
\catcode`\@=11
\def\psfortextures{
\def\PSspeci@l##1##2{%
\special{illustration ##1\space scaled ##2}%
}}%
\def\psfordvitops{
\def\PSspeci@l##1##2{%
\special{dvitops: import ##1\space \the\drawingwd \the\drawinght}%
}}%
\def\psfordvips{
\def\PSspeci@l##1##2{%
\d@my=0.1bp \d@mx=\drawingwd \divide\d@mx by\d@my
\includegraphics{##1\space}}}%
\def\psforoztex{
\def\PSspeci@l##1##2{%
\special{##1 \space
      ##2 1000 div dup scale
      \number-\psllx\space\space \number-\pslly\space\space translate
}}}%
\def\psfordvitps{
\def\dvitpsLiter@ldim##1{\dimen0=##1\relax
\special{dvitps: Literal "\number\dimen0\space"}}%
\def\PSspeci@l##1##2{%
\at(0bp;\drawinght){%
\special{dvitps: Include0 "psfig.psr"}
\dvitpsLiter@ldim{\drawingwd}%
\dvitpsLiter@ldim{\drawinght}%
\dvitpsLiter@ldim{\psllx bp}%
\dvitpsLiter@ldim{\pslly bp}%
\dvitpsLiter@ldim{\psurx bp}%
\dvitpsLiter@ldim{\psury bp}%
\special{dvitps: Literal "startTexFig"}%
\special{dvitps: Include1 "##1"}%
\special{dvitps: Literal "endTexFig"}%
}}}%
\def\psfordvialw{
\def\PSspeci@l##1##2{
\special{language "PostScript",
position = "bottom left",
literal "  \psllx\space \pslly\space translate
  ##2 1000 div dup scale
  -\psllx\space -\pslly\space translate",
include "##1"}
}}%
\def\psforptips{
\def\PSspeci@l##1##2{{
\d@mx=\psurx bp
\advance \d@mx by -\psllx bp
\divide \d@mx by 1000\multiply\d@mx by \xscale
\incm{\d@mx}
\let\tmpx\dimincm
\d@my=\psury bp
\advance \d@my by -\pslly bp
\divide \d@my by 1000\multiply\d@my by \xscale
\incm{\d@my}
\let\tmpy\dimincm
\d@mx=-\psllx bp
\divide \d@mx by 1000\multiply\d@mx by \xscale
\d@my=-\pslly bp
\divide \d@my by 1000\multiply\d@my by \xscale
\at(\d@mx;\d@my){\special{ps:##1 x=\tmpx cm, y=\tmpy cm}}
}}}%
\def\psonlyboxes{
\def\PSspeci@l##1##2{%
\at(0cm;0cm){\boxit{\vbox to\drawinght
  {\vss\hbox to\drawingwd{\at(0cm;0cm){\hbox{({\tt##1})}}\hss}}}}
}}%
\def\psloc@lerr#1{%
\let\savedPSspeci@l=\PSspeci@l%
\def\PSspeci@l##1##2{%
\at(0cm;0cm){\boxit{\vbox to\drawinght
  {\vss\hbox to\drawingwd{\at(0cm;0cm){\hbox{({\tt##1}) #1}}\hss}}}}
\let\PSspeci@l=\savedPSspeci@l
}}%
\newread\pst@mpin
\newdimen\drawinght\newdimen\drawingwd
\newdimen\psxoffset\newdimen\psyoffset
\newbox\drawingBox
\newcount\xscale \newcount\yscale \newdimen\pscm\pscm=1cm
\newdimen\d@mx \newdimen\d@my
\newdimen\pswdincr \newdimen\pshtincr
\let\ps@nnotation=\relax
{\catcode`\|=0 |catcode`|\=12 |catcode`|
|catcode`#=12 |catcode`*=14
|xdef|backslashother{\}*
|xdef|percentother{
|xdef|tildeother{~}*
|xdef|sharpother{#}*
}%
\def\R@moveMeaningHeader#1:->{}%
\def\uncatcode#1{%
\edef#1{\expandafter\R@moveMeaningHeader\meaning#1}}%
\def\execute#1{#1}
\def\psm@keother#1{\catcode`#112\relax}
\def\executeinspecs#1{%
\execute{\begingroup\let\do\psm@keother\dospecials\catcode`\^^M=9#1\endgroup}}%
\def\@mpty{}%
\def\matchexpin#1#2{
  \fi%
  \edef\tmpb{{#2}}%
  \expandafter\makem@tchtmp\tmpb%
  \edef\tmpa{#1}\edef\tmpb{#2}%
  \expandafter\expandafter\expandafter\m@tchtmp\expandafter\tmpa\tmpb\endm@tch%
  \if\match%
}%
\def\matchin#1#2{%
  \fi%
  \makem@tchtmp{#2}%
  \m@tchtmp#1#2\endm@tch%
  \if\match%
}%
\def\makem@tchtmp#1{\def\m@tchtmp##1#1##2\endm@tch{%
  \def\tmpa{##1}\def\tmpb{##2}\let\m@tchtmp=\relax%
  \ifx\tmpb\@mpty\def\match{YN}%
  \else\def\match{YY}\fi%
}}%
\def\incm#1{{\psxoffset=1cm\d@my=#1
 \d@mx=\d@my
  \divide\d@mx by \psxoffset
  \xdef\dimincm{\number\d@mx.}
  \advance\d@my by -\number\d@mx cm
  \multiply\d@my by 100
 \d@mx=\d@my
  \divide\d@mx by \psxoffset
  \edef\dimincm{\dimincm\number\d@mx}
  \advance\d@my by -\number\d@mx cm
  \multiply\d@my by 100
 \d@mx=\d@my
  \divide\d@mx by \psxoffset
  \xdef\dimincm{\dimincm\number\d@mx}
}}%
\newif\ifNotB@undingBox
\newhelp\PShelp{Proceed: you'll have a 5cm square blank box instead of
your graphics.}%
\def\s@tsize#1 #2 #3 #4\@ndsize{
  \def\psllx{#1}\def\pslly{#2}%
  \def\psurx{#3}\def\psury{#4}
  \ifx\psurx\@mpty\NotB@undingBoxtrue
  \else
    \drawinght=#4bp\advance\drawinght by-#2bp
    \drawingwd=#3bp\advance\drawingwd by-#1bp
  \fi
  }%
\def\sc@nBBline#1:#2\@ndBBline{\edef\p@rameter{#1}\edef\v@lue{#2}}%
\def\g@bblefirstblank#1#2:{\ifx#1 \else#1\fi#2}%
{\catcode`\%=12
\xdef\B@undingBox{
\def\ReadPSize#1{
 \readfilename#1\relax
 \let\PSfilename=\lastreadfilename
 \openin\pst@mpin=#1\relax
 \ifeof\pst@mpin \errhelp=\PShelp
   \errmessage{I haven't found your postscript file (\PSfilename)}%
   \psloc@lerr{was not found}%
   \s@tsize 0 0 142 142\@ndsize
   \closein\pst@mpin
 \else
   \if\matchexpin{\GlobalInputList}{, \lastreadfilename}%
   \else\xdef\GlobalInputList{\GlobalInputList, \lastreadfilename}%
     \immediate\write\psbj@inaux{\lastreadfilename,}%
   \fi%
   \loop
     \executeinspecs{\catcode`\ =10\global\read\pst@mpin to\n@xtline}%
     \ifeof\pst@mpin
       \errhelp=\PShelp
       \errmessage{(\PSfilename) is not an Encapsulated PostScript File:
           I could not find any \B@undingBox: line.}%
       \edef\v@lue{0 0 142 142:}%
       \psloc@lerr{is not an EPSFile}%
       \NotB@undingBoxfalse
     \else
       \expandafter\sc@nBBline\n@xtline:\@ndBBline
       \ifx\p@rameter\B@undingBox\NotB@undingBoxfalse
         \edef\t@mp{%
           \expandafter\g@bblefirstblank\v@lue\space\space\space}%
         \expandafter\s@tsize\t@mp\@ndsize
       \else\NotB@undingBoxtrue
       \fi
     \fi
   \ifNotB@undingBox\repeat
   \closein\pst@mpin
 \fi
\message{#1}%
}%
\def\psboxto(#1;#2)#3{\vbox{%
   \ReadPSize{#3}%
   \advance\pswdincr by \drawingwd
   \advance\pshtincr by \drawinght
   \divide\pswdincr by 1000
   \divide\pshtincr by 1000
   \d@mx=#1
   \ifdim\d@mx=0pt\xscale=1000
         \else \xscale=\d@mx \divide \xscale by \pswdincr\fi
   \d@my=#2
   \ifdim\d@my=0pt\yscale=1000
         \else \yscale=\d@my \divide \yscale by \pshtincr\fi
   \ifnum\yscale=1000
         \else\ifnum\xscale=1000\xscale=\yscale
                    \else\ifnum\yscale<\xscale\xscale=\yscale\fi
              \fi
   \fi
   \divide\drawingwd by1000 \multiply\drawingwd by\xscale
   \divide\drawinght by1000 \multiply\drawinght by\xscale
   \divide\psxoffset by1000 \multiply\psxoffset by\xscale
   \divide\psyoffset by1000 \multiply\psyoffset by\xscale
   \global\divide\pscm by 1000
   \global\multiply\pscm by\xscale
   \multiply\pswdincr by\xscale \multiply\pshtincr by\xscale
   \ifdim\d@mx=0pt\d@mx=\pswdincr\fi
   \ifdim\d@my=0pt\d@my=\pshtincr\fi
   \message{scaled \the\xscale}%
 \hbox to\d@mx{\hss\vbox to\d@my{\vss
   \global\setbox\drawingBox=\hbox to 0pt{\kern\psxoffset\vbox to 0pt{%
      \kern-\psyoffset
      \PSspeci@l{\PSfilename}{\the\xscale}%
      \vss}\hss\ps@nnotation}%
   \global\wd\drawingBox=\the\pswdincr
   \global\ht\drawingBox=\the\pshtincr
   \global\drawingwd=\pswdincr
   \global\drawinght=\pshtincr
   \baselineskip=0pt
   \copy\drawingBox
 \vss}\hss}%
  \global\psxoffset=0pt
  \global\psyoffset=0pt
  \global\pswdincr=0pt
  \global\pshtincr=0pt 
  \global\pscm=1cm 
}}%
\def\psboxscaled#1#2{\vbox{%
  \ReadPSize{#2}%
  \xscale=#1
  \message{scaled \the\xscale}%
  \divide\pswdincr by 1000 \multiply\pswdincr by \xscale
  \divide\pshtincr by 1000 \multiply\pshtincr by \xscale
  \divide\psxoffset by1000 \multiply\psxoffset by\xscale
  \divide\psyoffset by1000 \multiply\psyoffset by\xscale
  \divide\drawingwd by1000 \multiply\drawingwd by\xscale
  \divide\drawinght by1000 \multiply\drawinght by\xscale
  \global\divide\pscm by 1000
  \global\multiply\pscm by\xscale
  \global\setbox\drawingBox=\hbox to 0pt{\kern\psxoffset\vbox to 0pt{%
     \kern-\psyoffset
     \PSspeci@l{\PSfilename}{\the\xscale}%
     \vss}\hss\ps@nnotation}%
  \advance\pswdincr by \drawingwd
  \advance\pshtincr by \drawinght
  \global\wd\drawingBox=\the\pswdincr
  \global\ht\drawingBox=\the\pshtincr
  \global\drawingwd=\pswdincr
  \global\drawinght=\pshtincr
  \baselineskip=0pt
  \copy\drawingBox
  \global\psxoffset=0pt
  \global\psyoffset=0pt
  \global\pswdincr=0pt
  \global\pshtincr=0pt 
  \global\pscm=1cm
}}%
\def\psbox#1{\psboxscaled{1000}{#1}}%
\newif\ifn@teof\n@teoftrue
\newif\ifc@ntrolline
\newif\ifmatch
\newread\j@insplitin
\newwrite\j@insplitout
\newwrite\psbj@inaux
\immediate\openout\psbj@inaux=psbjoin.aux
\immediate\write\psbj@inaux{\string\joinfiles}%
\immediate\write\psbj@inaux{\jobname,}%
\def\toother#1{\ifcat\relax#1\else\expandafter%
  \toother@ux\meaning#1\endtoother@ux\fi}%
\def\toother@ux#1 #2#3\endtoother@ux{\def\tmp{#3}%
  \ifx\tmp\@mpty\def\tmp{#2}\let\next=\relax%
  \else\def\next{\toother@ux#2#3\endtoother@ux}\fi%
\next}%
\let\readfilenamehook=\relax
\def\re@d{\expandafter\re@daux}
\def\re@daux{\futurelet\nextchar\stopre@dtest}%
\def\re@dnext{\xdef\lastreadfilename{\lastreadfilename\nextchar}%
  \afterassignment\re@d\let\nextchar}%
\def\stopre@d{\egroup\readfilenamehook}%
\def\stopre@dtest{%
  \ifcat\nextchar\relax\let\nextread\stopre@d
  \else
    \ifcat\nextchar\space\def\nextread{%
      \afterassignment\stopre@d\chardef\nextchar=`}%
    \else\let\nextread=\re@dnext
      \toother\nextchar
      \edef\nextchar{\tmp}%
    \fi
  \fi\nextread}%
\def\readfilename{\bgroup%
  \let\\=\backslashother \let\%=\percentother \let\~=\tildeother
  \let\#=\sharpother \xdef\lastreadfilename{}%
  \re@d}%
\xdef\GlobalInputList{\jobname}%
\def\psnewinput{%
  \def\readfilenamehook{
    \if\matchexpin{\GlobalInputList}{, \lastreadfilename}%
    \else\xdef\GlobalInputList{\GlobalInputList, \lastreadfilename}%
      \immediate\write\psbj@inaux{\lastreadfilename,}%
    \fi%
    \let\readfilenamehook=\relax%
    \ps@ldinput\lastreadfilename\relax%
  }\readfilename%
}%
\expandafter\ifx\csname @@input\endcsname\relax    
  \immediate\let\ps@ldinput=\input\def\input{\psnewinput}%
\else
  \immediate\let\ps@ldinput=\@@input
  \def\@@input{\psnewinput}%
\fi%
\def\nowarnopenout{%
 \def\warnopenout##1##2{%
   \readfilename##2\relax
   \message{\lastreadfilename}%
   \immediate\openout##1=\lastreadfilename\relax}}%
\def\warnopenout#1#2{%
 \readfilename#2\relax
 \def\t@mp{TrashMe,psbjoin.aux,psbjoint.tex,}\uncatcode\t@mp
 \if\matchexpin{\t@mp}{\lastreadfilename,}%
 \else
   \immediate\openin\pst@mpin=\lastreadfilename\relax
   \ifeof\pst@mpin
     \else
     \edef\tmp{{If the content of this file is precious to you, this
is your last chance to abort (ie press x or e) and rename it before
retexing (\jobname). If you're sure there's no file
(\lastreadfilename) in the directory of (\jobname), then go on: I'm
simply worried because you have another (\lastreadfilename) in some
directory I'm looking in for inputs...}}%
     \errhelp=\tmp
     \errmessage{I may be about to replace your file named \lastreadfilename}%
   \fi
   \immediate\closein\pst@mpin
 \fi
 \message{\lastreadfilename}%
 \immediate\openout#1=\lastreadfilename\relax}%
{\catcode`\%=12\catcode`\*=14
\gdef\splitfile#1{*
 \readfilename#1\relax
 \immediate\openin\j@insplitin=\lastreadfilename\relax
 \ifeof\j@insplitin
   \message{! I couldn't find and split \lastreadfilename!}*
 \else
   \immediate\openout\j@insplitout=TrashMe
   \message{< Splitting \lastreadfilename\space into}*
   \loop
     \ifeof\j@insplitin
       \immediate\closein\j@insplitin\n@teoffalse
     \else
       \n@teoftrue
       \executeinspecs{\global\read\j@insplitin to\spl@tinline\expandafter
         \ch@ckbeginnewfile\spl@tinline
       \ifc@ntrolline
       \else
         \toks0=\expandafter{\spl@tinline}*
         \immediate\write\j@insplitout{\the\toks0}*
       \fi
     \fi
   \ifn@teof\repeat
   \immediate\closeout\j@insplitout
 \fi\message{>}*
}*
\gdef\ch@ckbeginnewfile#1
 \def\t@mp{#1}*
 \ifx\@mpty\t@mp
   \def\t@mp{#3}*
   \ifx\@mpty\t@mp
     \global\c@ntrollinefalse
   \else
     \immediate\closeout\j@insplitout
     \warnopenout\j@insplitout{#2}*
     \global\c@ntrollinetrue
   \fi
 \else
   \global\c@ntrollinefalse
 \fi}*
\gdef\joinfiles#1\into#2{*
 \message{< Joining following files into}*
 \warnopenout\j@insplitout{#2}*
 \message{:}*
 {*
 \edef\w@##1{\immediate\write\j@insplitout{##1}}*
\w@{
\w@{
\w@{
\w@{
\w@{
\w@{
\w@{
\w@{
\w@{
\w@{
\w@{\string\input\space psbox.tex}*
\w@{\string\splitfile{\string\jobname}}*
\w@{\string\let\string\autojoin=\string\relax}*
}*
 \expandafter\tre@tfilelist#1, \endtre@t
 \immediate\closeout\j@insplitout
 \message{>}*
}*
\gdef\tre@tfilelist#1, #2\endtre@t{*
 \readfilename#1\relax
 \ifx\@mpty\lastreadfilename
 \else
   \immediate\openin\j@insplitin=\lastreadfilename\relax
   \ifeof\j@insplitin
     \errmessage{I couldn't find file \lastreadfilename}*
   \else
     \message{\lastreadfilename}*
     \immediate\write\j@insplitout{
     \executeinspecs{\global\read\j@insplitin to\oldj@ininline}*
     \loop
       \ifeof\j@insplitin\immediate\closein\j@insplitin\n@teoffalse
       \else\n@teoftrue
         \executeinspecs{\global\read\j@insplitin to\j@ininline}*
         \toks0=\expandafter{\oldj@ininline}*
         \let\oldj@ininline=\j@ininline
         \immediate\write\j@insplitout{\the\toks0}*
       \fi
     \ifn@teof
     \repeat
   \immediate\closein\j@insplitin
   \fi
   \tre@tfilelist#2, \endtre@t
 \fi}*
}%
\def\autojoin{%
 \immediate\write\psbj@inaux{\string\into{psbjoint.tex}}%
 \immediate\closeout\psbj@inaux
 \expandafter\joinfiles\GlobalInputList\into{psbjoint.tex}%
}%
\def\centinsert#1{\midinsert\line{\hss#1\hss}\endinsert}%
\def\psannotate#1#2{\vbox{%
  \def\ps@nnotation{#2\global\let\ps@nnotation=\relax}#1}}%
\def\pscaption#1#2{\vbox{%
   \setbox\drawingBox=#1
   \copy\drawingBox
   \vskip\baselineskip
   \vbox{\hsize=\wd\drawingBox\setbox0=\hbox{#2}%
     \ifdim\wd0>\hsize
       \noindent\unhbox0\tolerance=5000
    \else\centerline{\box0}%
    \fi
}}}%
\def\at(#1;#2)#3{\setbox0=\hbox{#3}\ht0=0pt\dp0=0pt
  \rlap{\kern#1\vbox to0pt{\kern-#2\box0\vss}}}%
\newdimen\gridht \newdimen\gridwd
\def\gridfill(#1;#2){%
  \setbox0=\hbox to 1\pscm
  {\vrule height1\pscm width.4pt\leaders\hrule\hfill}%
  \gridht=#1
  \divide\gridht by \ht0
  \multiply\gridht by \ht0
  \gridwd=#2
  \divide\gridwd by \wd0
  \multiply\gridwd by \wd0
  \advance \gridwd by \wd0
  \vbox to \gridht{\leaders\hbox to\gridwd{\leaders\box0\hfill}\vfill}}%
\def\fillinggrid{\at(0cm;0cm){\vbox{%
  \gridfill(\drawinght;\drawingwd)}}}%
\def\textleftof#1:{%
  \setbox1=#1
  \setbox0=\vbox\bgroup
    \advance\hsize by -\wd1 \advance\hsize by -2em}%
\def\textrightof#1:{%
  \setbox0=#1
  \setbox1=\vbox\bgroup
    \advance\hsize by -\wd0 \advance\hsize by -2em}%
\def\endtext{%
  \egroup
  \hbox to \hsize{\valign{\vfil##\vfil\cr%
\box0\cr%
\noalign{\hss}\box1\cr}}}%
\def\frameit#1#2#3{\hbox{\vrule width#1\vbox{%
  \hrule height#1\vskip#2\hbox{\hskip#2\vbox{#3}\hskip#2}%
        \vskip#2\hrule height#1}\vrule width#1}}%
\def\boxit#1{\frameit{0.4pt}{0pt}{#1}}%
\catcode`\@=12 
\psfordvips   

\usepackage{graphics}
\usepackage{multirow}
\usepackage[usenames]{color}
\usepackage{nicefrac}
\usepackage{amsmath}

\newcommand {\mb}[1]{\mbox{\boldmath{${#1}$}}}

\begin{document}

\title{Prototyping method for Bragg--type atom interferometers}
\author{Brandon Benton}
\affiliation{Department of Physics, Georgia Southern University,
Statesboro, GA 30460--8031 USA}
\author{Michael Krygier}
\affiliation{Department of Physics, Georgia Southern University,
Statesboro, GA 30460--8031 USA}
\author{Jeffrey Heward}
\affiliation{Department of Physics, Georgia Southern University,
Statesboro, GA 30460--8031 USA}
\author{Mark Edwards}
\affiliation{Department of Physics, Georgia Southern University,
Statesboro, GA 30460--8031 USA}
\author{Charles W.\ Clark}
\affiliation{Joint Quantum Insitute, National Institute of Standards 
and Technology and the University of Maryland, Gaithersburg, MD 20899, USA}

\date{\today}

\begin{abstract}
We present a method for {\em rapid} modeling of new Bragg 
ultra--cold atom interferometer (AI) designs useful for assessing the per%
formance of such interferometers. The method simulates the overall effect 
on the condensate wave function in a given AI design using two separate 
elements.  These are (1) modeling the effect of a Bragg pulse on the wave 
function and (2) approximating the evolution of the wave function during
the intervals between the pulses. The actual sequence of these pulses and 
intervals is then followed to determine the approximate final wave function 
from which the interference pattern can be calculated.  The exact evolution 
between pulses is assumed to be governed by the Gross--Pitaevskii (GP) equation 
whose solution is approximated using a Lagrangian Variational Method to 
facilitate rapid estimation of performance.  The method
presented here is an extension of an earlier one that was used to analyze 
the results of an experiment [J.E.\ Simsarian, et al., \prl {\bf 83}, 2040
(2000)], where the phase of a Bose--Einstein condensate was measured using 
a Mach--Zehnder--type Bragg AI.  We have developed both 1D and 3D versions 
of this method and we have determined their validity by comparing their
predicted interference patterns with those obtained by numerical integration 
of the 1D GP equation and with the results of the above experiment.  We find
excellent agreement between the 1D interference patterns predicted by this
method and those found by the GP equation.  We show that we can reproduce 
all of the results of that experiment without recourse to an {\em ad hoc}
velocity--kick correction needed by the earlier method, including some
experimental results that the earlier model did not predict.  We also found 
that this method provides estimates of 1D interference patterns at least four
orders--of--magnitude faster than direct numerical solution of the 1D GP
equation.
\end{abstract}

\pacs{03.75.Dg,67.85.Hj,03.67.Lx,03.75.Kk,42.50.Gy}

\maketitle

\section{Introduction}
\label{intro}

It is possible to use ideas inspired by advances in quantum information
science (QIS) to devise improved--performance matter--wave interferometers.  
A recent example of this has been seen in neutron interferometry where 
the idea of decoherence--free subspaces was used to redesign a neutron interferometer to reduce the effect of mechanical shaking on the interference
contrast~\cite{PhysRevA.79.053635,PhysRevLett.107.150401}.  The use of ideas
from QIS to drive new neutron interferometer designs may also be possible 
for atom interferometers. Two promising areas of QIS where this could happen
include decoherence avoidance and minimization.  Interferometer applications
where QIS--inspired redesigns may result in improved performance include precision navigation and metrology.  This paper presents a tool for rapid
assessment of new atom-interferometer designs for such applications.

Atom interferometers (AI), where laser light is applied to ultra--cold 
atoms, have many applications.  These include  
quantum decoherence~\cite{PhysRevLett.91.090408,AngewChemIntEd.47.6195}, 
properties of Bose--Einstein condensates~\cite{PhysRevLett.78.582,PhysRev%
Lett.82.3008,PhysRevLett.83.5407,Kozuma17121999}, precision measurement 
of the fine-structure constant~\cite{PhysRevLett.106.080801,2011arXiv1103%
.1454J} and testing the charge neutrality of atoms~\cite{PhysRevA.47.4663}. 
Atom interferometers are also used in many precision measurement devices.  
These include gravimeters, gyroscopes, and gradiometers which all have 
important applications in precision navigation~\cite{Metrologia.38.25,%
PhysRevA.65.033608,ClassQuantumGrav.17.2385}.  Atom interferometers also 
have applications in atomic physics such as atomic polarizability measure%
ments and Casimir--Polder potentials for atoms near surfaces~\cite{EurPhys%
JD.38.353}.  More uses of atom interferometry are described in Ref.\ \cite
{RevModPhys.81.1051}.

With the advent of gaseous Bose--Einstein condensates (BEC)~\cite{%
Science.269.198,PhysRevLett.75.1687,PhysRevLett.75.3969,Pethick_and_%
Smith,Pitaevskii_and_Stringari}, strong interest has developed in 
using AIs for precision metrology~\cite{PhysRevLett.89.140401,PhysRevA%
.81.043633,PhysRevLett.98.200801,PhysRevA.78.023619,arXiv:0505358,%
PhysRevA.77.043604,2011arXiv1103.1454J,arXiv:1011.5804}.  Most of these
ultra--cold atom interferometers were of the standard Mach--Zehnder design.
However, some more recent precision interferometers~\cite{2011arXiv1103.%
1454J} have different designs.  This also suggests that
advances in interferometer design may lead to significant AI performance 
gains.  

There are many factors that can limit the performance of an AI.  Some
of these include mirror vibration, random initial motion of BECs at 
birth, stray light, external magnetic fields, and errors in the frequency
or intensity of the applied laser light.  Atom interferometers confined
on an atom chip can have other problems related to atom loss, heating,
and decoherence~\cite{FolmanAdvAtMolOptPhys}.  One of the ways in which
some of these factors may be addressed is with new AI designs.

In order to pursue the program of drawing ideas from QIS to inspire new
AI designs, it will be necessary to develop tools that can be used to provide
{\em rapid} assessment of the performance of these new designs. In this work 
we present a method for rapid simulation of the behavior of condensates in 
Bragg interferometers.  We assume that the evolution of the condensate be%
tween Bragg pulses is described by the Gross--Pitaevskii (GP) equation. The
method approximates the evolution of the condensate wave function by modeling
the effect of individual pulses and its evolution between pulses.  Thus the
final condensate wave function can be found enabling the prediction of the 
final interference pattern.  As will be seen below, our method provides rea%
sonable estimates of AI behavior in a time that is four orders--of--magnitude
faster than that needed for numerical solution of the (1D) GP equation to
simulate a standard Mach--Zehnder AI.  The time--savings factor for 3D sim%
ulation will be greater.  Such a tool will be essential for preliminary test%
ing of new AI designs having more pulses and longer evolution times.

In Section \ref{bragg_interferometer} we describe the standard $\pi/2$--%
$\pi$--$\pi/2$ Bragg interferometer followed in Section \ref{bragg_models}
by a general overview of the two elements of our Bragg prototyper model
and the details of the original model used in the analysis of a Bragg inter%
ferometer experiment carried out at NIST~\cite{PhysRevLett.85.2040}. In
the original model each definite--momentum cloud in the condensate wave 
function was represented by a single gaussian.  An extra velocity ``kick'', 
caused by the repulsion of the separating clouds after a $\pi/2$ Bragg 
pulse, was needed to obtain agreement with the experimental results.  A 
study of this kick for a single pulse was conducted as a function of 
interaction strength of the condensate and the results are presented in 
Section \ref{repulsion_study}.  Section \ref{two_cloud} presents the
two--cloud version of our Bragg prototyper model for both the 1D and 3D
cases.  Finally, Section \ref{discussion} contains a summary and discussion
of the possible applications of the method.

\section{Bragg atom interferometer}
\label{bragg_interferometer}

A Bose--Einstein condensate (BEC) can be coherently split into two
clouds, a fast--moving cloud and a slow--moving cloud, through the 
application of a Bragg pulse~\cite{PhysRevLett.82.871}.  
If the condensate is stationary when a Bragg pulse is applied to it, then
the result will be two clouds, one that remains stationary and another
whose momentum is $\hbar\Delta{\bf k}=\hbar({\bf k}_{1}-{\bf k}_{2})$ 
where ${\bf k}_{1}$ (${\bf k}_{2}$) is the photon momentum of the higher 
(lower) frequency laser beam.  If the condensate is moving with this 
momentum when the Bragg pulse is applied, then the result is again two 
clouds one of which keeps its momentum while the other cloud's momentum 
is reduced by $\hbar\Delta{\bf k}$.  In either case, the net result of
applying a Bragg pulse to a condensate is a fast cloud and a slow cloud. 
One of these clouds has the momentum of the original cloud and the 
momentum of the other cloud is increased (decreased) if the original cloud 
was slow (fast).  

\begin{figure}[htb]
\begin{center}
\mbox{\psboxto(3.2in;0.0in){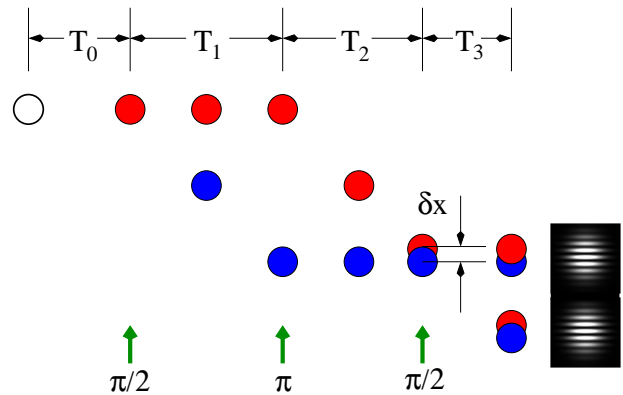}}
\end{center}
\caption{(color online) This figure shows the sequence of Bragg 
pulses (represented by green arrows) applied to a condensate in a
$\pi/2$--$\pi$--$\pi/2$ Bragg interferometer.  The vertical (horizontal)
direction represents space (time).  See text for further details.}
\label{bragg_mz_fig}
\end{figure} 

A Mach--Zehnder--type Bragg interferometer can be constructed by applying
three Bragg pulses in the sequence $\pi/2$--$\pi$--$\pi/2$ with variable
time intervals between them~\cite{PhysRevA.61.041602}. This is shown in 
Fig.\ \ref{bragg_mz_fig} where the open circle shows the initial
condensate which may be released from the trap and allowed to expand
for a time $T_{0}$.  The first $\pi/2$ pulse splits the condensate into 
a slow cloud (upper path) and a fast cloud (lower path).  After a time
interval $T_{1}$ the $\pi$ pulse is applied which stops the fast (lower) 
cloud and starts the slow cloud so that the two clouds come back together. 
The final pulse is applied after a time $T_{2}$ when the two clouds again 
overlap causing each cloud to split again.  After a further time $T_{3}$
there is a pair of overlapping slow clouds and a pair of overlapping 
fast clouds that give rise to the final pattern that is imaged at that
moment.

\section{Bragg interferometer prototyper models}
\label{bragg_models}

The operation of an arbitrary Bragg AI can be specified by stating the
times and angles of the Bragg pulses applied to the condensate.  Although
it is possible to apply a Bragg pulse having an arbitrary angle, $\theta$,
here we will restrict our attention to pulses where $\theta=\pi$ or
$\theta=\pi/2$ radians.  Hence these pulses will either split clouds 
into two equal pieces, one fast and one slow, or will leave the condensate
whole and merely swap its fast or slow velocity.  Thus we only consider
interferometer sequences that are composed of $\pi/2$ and/or $\pi$ pulses.

\subsection{Overview of prototyper models}
\label{models_overview}

Our general prototyper model will enable us to approximate the evolution 
of the condensate wave function through all of the steps of any given 
interferometer sequence.  We assume that the duration of all Bragg pulses
is short compared to the characteristic time for collective effects of
the condensate to be manifested.  This time is usually ~$\hbar/\mu$ where
$\mu$ is the chemical potential of the condensate.  This means that we
assume that any changes in the momentum space distribution of the condensate
atoms caused by a Bragg pulse happen instantaneously.  We further assume 
that the characteristic size of any momentum change is large in the sense
that the wavelength of the photon causing the change is small compared to
the size of the condensate.  Finally we assume that the evolution of the
condensate between pulses is governed by the Gross--Pitaevskii equation.

Our Bragg prototyper method therefore has two essential elements: (1) an 
approximation of the effect of a Bragg pulse on the condensate wave 
function, and (2) a model for approximating the GP--governed condensate
wave function behavior between pulses.  The method consists of applying
these two elements to the pulse sequence of the given interferometer design 
to produce an approximate final condensate wave function so that predictions 
about the measured interference patterns can be made.

For the first element, as will be described in more detail below, we represent 
the condensate wave function at any moment as a superposition of a number
of fast or slow gaussian clouds.  We model the effect of Bragg pulses as
follows.  If the wave function for a given cloud in the condensate wave
function is $\psi({\bf r}, t)$ before the Bragg pulse, then the change in 
this wave function is, for\\{\bf slow clouds}:
\begin{eqnarray}
\psi({\bf r},t) 
&\rightarrow&
\frac{1}{\sqrt{2}}
\left(\psi({\bf r},t) - 
e^{-i\phi}e^{i\Delta{\bf k}\cdot{\bf r}}\psi({\bf r},t)
\right)
\quad
(\pi/2\ {\rm pulse})\nonumber\\
\psi({\bf r},t)
&\rightarrow&
-e^{-i\phi}e^{i\Delta{\bf k}\cdot{\bf r}}\psi({\bf r},t)
\quad
(\pi\ {\rm pulse})
\label{slow_clouds}
\end{eqnarray}
and for {\bf fast clouds}:
\begin{eqnarray}
\psi({\bf r},t) 
&\rightarrow&
\frac{1}{\sqrt{2}}
\left(e^{i\phi}\psi({\bf r},t) + 
e^{-i\Delta{\bf k}\cdot{\bf r}}\psi({\bf r},t)
\right)
\quad
(\pi/2\ {\rm pulse})\nonumber\\
\psi({\bf r},t)
&\rightarrow&
e^{i\phi}e^{-i\Delta{\bf k}\cdot{\bf r}}\psi({\bf r},t)
\quad
(\pi\ {\rm pulse})
\label{fast_clouds}
\end{eqnarray}
where $\Delta{\bf k}$ is the momentum change for Bragg pulses defined 
above.  The factor $\phi$ is the phase of the moving standing wave
in the center of the initial atomic wavepacket in the middle
of the Bragg pulse~\cite{PhysRevA.61.041602}.  Thus the action of a $\pi/2$
pulse is to double the number of clouds, adding a new fast cloud on top 
of a previously existing slow cloud and adding a new slow cloud on top of a
previously existing fast cloud.  In each case the shape of the previously
existing cloud is unchanged.  The action of a $\pi$ pulse is to convert a
previously existing slow (fast) cloud into a fast (slow) cloud.

For the second element we use the Lagrangian Variational Method (LVM)~\cite
{PhysRevA.56.1424,0953-4075-38-4-004} to approximate the condensate evolution 
between pulses. Although it is possible to solve the 3D GP equation to 
determine this evolution, this is not practical for {\em rapid} estimation
of the final condensate wave function.  The LVM provides approximate
solutions to the GP equation in the form of equations of motion for
time--dependent parameters that appear in an assumed trial wave function.
Thus the exact solution of the GP equation that requires the solution of
a 3+1 partial differential equation is traded for approximate solutions
that can be obtained by solving a system of ordinary differential equations
in time.  We briefly review this method now.

The GP equation is given by
\begin{equation}
i\hbar\frac{\partial\Psi}{\partial t} = 
-\frac{\hbar^{2}}{2m}\nabla^{2}\Psi + V_{\rm trap}({\bf r})\Psi
+ gN\left|\Psi\right|^{2}\Psi
\end{equation}
where we assume that any trapping potential is harmonic:
\begin{equation}
V_{\rm trap}({\bf r}) = 
\tfrac{1}{2}m\omega_{x}^{2}x^{2} +
\tfrac{1}{2}m\omega_{y}^{2}y^{2} +
\tfrac{1}{2}m\omega_{z}^{2}z^{2}.
\end{equation}
where $m$ is the mass of a condensate atom.

Here we will introduce scaled variables that will be used throughout the
rest of the paper.  First we choose a length unit appropriate to the 
harmonic potential: $L_{0}=(\hbar/2m\bar{\omega})^{1/2}$ where 
$\bar{\omega}=(\omega_{x}\omega_{y}\omega_{z})^{1/3}$ and then define the
energy unit as $E_{0}=\hbar^{2}/2mL_{0}^{2}$ and the time unit as
$T_{0}=\hbar/E_{0}$.  We then introduce scaled position and time variables
as $\bar{x}=x/L_{0},\bar{y}=y/L_{0},\bar{z}=z/L_{0},\bar{t}=t/T_{0}$, and
use barred quantities in general to represent quantities expressed in the
scaled units.  If, additionally, we write the condensate wave function in
scaled units as $\Psi=\bar{\Psi}/L_{0}^{3/2}$ then the GP equation becomes
\begin{equation}
i\frac{\partial\bar{\Psi}}{\partial\bar{t}} = 
-\bar{\nabla}^{2}\bar{\Psi} + \bar{V}_{\rm trap}({\bf \bar{r}})\bar{\Psi}
+ \bar{g}N\left|\bar{\Psi}\right|^{2}\bar{\Psi}
\end{equation}
where 
$\bar{\nabla}^{2}=
 \partial^{2}/\partial\bar{x}^{2}+
 \partial^{2}/\partial\bar{y}^{2}+
 \partial^{2}/\partial\bar{z}^{2}$ and 
\begin{equation}
\bar{V}_{\rm trap}({\bf \bar{r}}) = 
\tfrac{1}{4}\gamma_{x}^{2}\bar{x}^{2} +
\tfrac{1}{4}\gamma_{y}^{2}\bar{y}^{2} +
\tfrac{1}{4}\gamma_{z}^{2}\bar{z}^{2}
\end{equation}
and where $\gamma_{\eta}=\omega_{\eta}/\bar{w},\eta=x,y,z$ and
$\bar{g}=g/(E_{0}L_{0}^{3})$.

The LVM produces equations of motion for the $m$ time--dependent variational
parameters that appear in a given trial wave function $\bar{\psi}_{\rm
 trial}({\bf \bar{r}};q_{1}(t),\dots,q_{m}(t))$.  The equations of motion
are obtained from the LVM Lagrangian via the usual Euler--Lagrange equations
of motion:
\begin{equation}
\frac{d}{dt}
\left(\frac{\partial L_{\rm LVM}}{\partial\dot{q}_{j}}\right) -
\frac{\partial L_{\rm LVM}}{\partial q_{j}} = 0.
\quad
j = 1,\dots,m.
\label{E_L_eqs}
\end{equation}
The LVM Lagrangian, in turn, is computed by integrating the LVM Lagrangian
density
\begin{equation}
L_{\rm LVM}(q_{1}(t),\dots,q_{m}(t)) = 
\int d^{3}r\,{\cal L}\left[\bar{\psi}_{\rm trial}({\bf\bar{r}},t)\right].
\label{lvm_lag}
\end{equation}
Finally, the LVM Lagrangian density that corresponds to the GP equation is
given by
\begin{eqnarray}
{\cal L}\left[\psi\right] 
&=& 
\tfrac{i}{2}
\left(\psi\psi_{t}^{\ast} - \psi^{\ast}\psi_{t}\right) +
\bar{\bf\nabla}\psi^{\ast}\cdot\bar{\bf\nabla}\psi + 
\bar{V}_{\rm trap}({\bf\bar{r}})\left|\psi\right|^{2}\nonumber\\
&+&
\tfrac{1}{2}\bar{g}N\left|\psi\right|^{4},
\label{lvm_lag_den}
\end{eqnarray}
where $\psi_{t}$ denotes the partial derivative of $\psi$ with respect to
$t$. In order to get equations of motion relevant for a Bragg interferometer,
we must choose a trial wave function.  In this work we present equations
of motion for single--cloud gaussian trial wave functions in three 
dimensions and two--cloud gaussian trial wave functions for both one and 
three dimensions.

An alternative approach to efficient approximate solution of the GP equation in
interferometric applications has recently been presented by Jamison {\em et al.}
\cite{2011arXiv1103.1454J}.  It is based on a generalization of the Thomas-Fermi
method originally proposed by Castin and Dum. \cite{PhysRevLett.77.5315}

\subsection{3D single--cloud LVM model}
\label{single_cloud}

In the single--cloud LVM model we choose the trial wave function to have
the form of a single, three--dimensional gaussian wavepacket, as was done
previously in Ref.\ \cite{PhysRevA.56.1424}
\begin{equation}
\bar{\Psi}({\bf \bar{r}},\bar{t}) = 
\bar{A}(t)\prod_{\eta=x,y,z}
e^{-(\bar{\eta}-\bar{\eta}_{0}(\bar{t}))^{2}/2\bar{w}_{\eta}^{2}(\bar{t})+
i\bar{\alpha}_{\eta}(\bar{t})\bar{\eta}+
i\bar{\beta}_{\eta}(\bar{t})\bar{\eta}^{2}}.
\label{single_cloud_wf}
\end{equation}
Here $(\bar{x}_{0},\bar{y}_{0},\bar{z}_{0})$ are the coordinates of the 
center of the wavepacket.  The quantities $(\bar{\alpha}_{x}, 
\bar{\alpha}_{y}, \bar{\alpha}_{z})$ are the linear phase coefficients 
which govern the motion of the wavepacket center.  The 
$(\bar{w}_{x},\bar{w}_{y},\bar{w}_{z})$ are the widths of the gaussian 
along the three axes and the $(\bar{\beta}_{x},\bar{\beta}_{y},
\bar{\beta}_{z})$ are the quadratic phase coefficients and govern the 
evolution of the widths.  Finally, $\bar{A}(t)$ is a normalization 
coefficient that will be removed from the Lagrangian later when the
normalization constraint is imposed.  We can take $\bar{A}$ to be real 
because, if $\bar{A}$ had a phase, it would represent an overall wave 
function phase which would not be physical.

\begin{figure}[t]
\begin{center}
\mbox{\psboxto(3.2in;0.0in){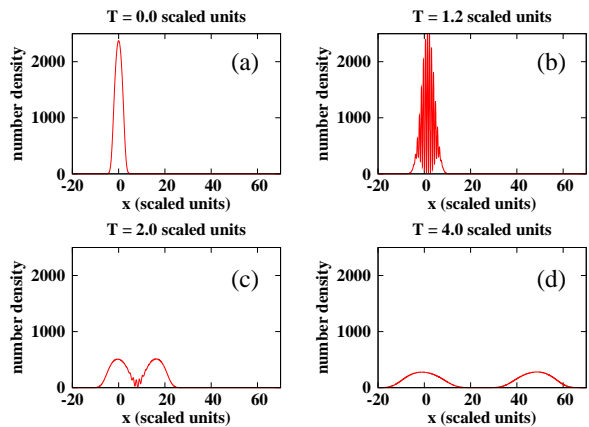}}
\end{center}
\caption{(color online) This figure shows the evolution of a 1D condensate
that is initially released and allowed to expand until $T=1$ scaled time 
unit at which time a $\pi/2$ Bragg pulse is applied.  The scaled velocity
imparted by the Bragg pulse to the fast cloud  is $\bar{v}_{L}=16$ scaled
velocity units.  (a) The initial condensate density just after the trap is
turned off, (b) after the $\pi/2$ Bragg pulse is applied, the density 
exhibits rapid oscillations in the region where the fast and slow clouds
overlap, (c) and (d) nearly complete and complete separation.  The value
of $\bar{g}N=10$ in scaled units.}
\label{bp_evol}
\end{figure} 
The equations of motion are derived as described above and the result 
is~\cite{PhysRevA.56.1424}
\begin{eqnarray}
\ddot{\bar{\eta}}_{0}+\gamma_{\eta}^{2}\bar{\eta}_{0} 
&=& 0\nonumber\\
\ddot{\bar{w}}_{\eta}+\gamma_{\eta}^{2}\bar{w}_{\eta}
&=&
\tfrac{4}{\bar{w}_{\eta}^{3}} +
\tfrac{2\bar{g}N}
{(2\pi)^{3/2}\bar{w}_{x}\bar{w}_{y}\bar{w}_{z}\bar{w}_{\eta}}\nonumber\\
\bar{\beta}_{\eta} 
&=& 
\tfrac{\dot{\bar{w}}_{\eta}}{4\bar{w}_{\eta}}\nonumber\\
\bar{\alpha}_{\eta}
&=&
\tfrac{1}{2}\dot{\bar{\eta}}_{0} - 2\bar{\beta}_{\eta}\bar{\eta}_{0}
\quad
\eta = x,y,z.
\end{eqnarray}
It is worth noting that our equations of motion differ slightly from 
those in Ref.~\cite{PhysRevA.56.1424} because of differences in the
definition of scaled units.

\subsection{NIST experiment and the single--cloud LVM model}
\label{nist_exp}

A Bragg AI was implemented at NIST and used to image the phase evolution 
of an evolving Bose--Einstein condensate~\cite{PhysRevLett.85.2040}.  In 
that experiment, a condensate of about $1.8\times10^{6}$ sodium atoms was 
held in a magnetic trap with trapping frequencies $\omega_{x}=\sqrt{2}
\omega_{y}=2\omega_{z}=2\pi\times27$ Hz and subjected to $\pi/2$--$\pi$--%
$\pi/2$ pulse sequence.  This sequence was performed both with the trap
left on and with the trap turned off and the condensate allowed to expand
for a time $T_{0}$.  With the trap on, the time conditions were fixed
so that $T_{1}=T_{2}\equiv T$ where $T$ was typically 1-2 ms and the clouds
were allowed to expand in order for them to separate before imaging.  In 
the trap--off case, a series of runs was carried out in which $T_{1}$ was held 
fixed at 1 ms while $T_{2}$ was varied such that the cloud overlap at the 
time of the final Bragg pulse ranged from fast cloud just arriving at the 
slow cloud until it had passed through and was just leaving.  This series 
of runs was performed for an expansion time of $T_{0}=1$ ms and repeated 
for $T_{0}=4$ ms.

The results of this experiment were analyzed using the single--cloud LVM 
just described~\cite{PhysRevLett.85.2040}.  While the results of the
single--cloud LVM model agreed well with experiment for runs performed
with the trap on, it did not agree with experiment for trap--off cases.  
This discrepancy was due to the presence of an extra relative velocity 
between interfering clouds at the moment of the final Bragg pulse.  The
extra velocity was caused by repulsion between overlapping clouds which 
occurred just after the first $\pi/2$ pulse and again just before the 
second $\pi/2$ pulse.  Agreement between theory and experiment was achieved 
by adding in by hand a small relative velocity correction to the condensate 
wave function predicted by the single--cloud model.  It is clear that any 
model able to account for this repulsion would need to include at least two
clouds.  In order to derive a simple model we need to validate a key
approximation by studying this correction for a single $\pi/2$ Bragg 
pulse.

\section{Repulsion study for single $\pi/2$ Bragg pulse}
\label{repulsion_study}

Before turning to a two--cloud model, we studied the effects of fast/slow 
cloud repulsion on the final relative velocity of the separating clouds after 
a single $\pi/2$ Bragg pulse.  We performed this study by simulating the
application of such a pulse on a 1D condensate and its subsequent evolution 
by numerical solution of the GP equation.  

In what follows we shall give the value of quantities in 1D scaled units 
which is a special case of the 3D scaled units given above.  In these units, 
all quantities are defined in terms of the length unit $L_{0}=\sqrt{\hbar
/2m\omega_{0}}$ which, in turn, is tied to a reference frequency, $\omega_{0}$.
This frequency is often the trap frequency but need not be as in the case
where the trap has been turned off.  Since these quantities are scaled out 
of the problem, it will be useful to give a numerical example of the sizes 
of the scaled units.  Thus, given a quasi--1D $^{87}$Rb condensate confined 
in a $\omega=2\pi\times10$ Hz trap, the length unit is $L_{0}=2.4\,\mu$m, 
the time unit is $T_{0}=15.8$ ms, and the velocity unit $v_{0}=0.015$ cm/s.
The value of the interaction strength is varied over the range $0\le\bar{g}
_{1D}N\le 200$ so that the transition from non--interacting up to the
Thomas--Fermi regime could be studied.  Here, $N$ is the number of condensate
atoms.

Figure \ref{bp_evol} shows a typical simulation where a condensate is 
released from the trap (panel (a)) and allowed to expand for until $T=1$ 
scaled time unit at which time a $\pi/2$ Bragg pulse is applied splitting 
the condensate into fast (on the right) and slow clouds.  During the 
separation the two clouds push each other apart so that the fast cloud moves
with a velocity $\bar{v}_{f}=\bar{v}_{L}+\delta\bar{v}$ that is slightly 
larger than the recoil velocity $\bar{v}_{L}$ caused by the laser light and 
the slow cloud drifts backwards with velocity $-\delta\bar{v}$.  In SI units,
the recoil velocity is $v_{L}=\hbar\Delta k/m$ and, in scaled units it can 
be expressed as $\bar{v}_{L}=v_{L}/(L_{0}/T_{0})=2\Delta\bar{k}$. The 
interaction strength of the initial condensate in this example was 
$\bar{g}N=10$ in scaled units.

To study this velocity ``kick'', $\delta\bar{v}$, we used the GP simulations to
determine its value as a function of the interaction strength of the initial
condensate.  The value of $\delta\bar{v}$ for a given value of $\bar{g}N$
was obtained by running a simulation where a $\pi/2$ Bragg pulse was applied. 
The velocities of the fast and slow clouds were determined, for each value
of $\bar{g}N$, as follows. In each run the two clouds were allow to separate
fully after the pulse; a midpoint between the two clouds, $\bar{x}_{\rm mid}$, 
and a time after which both clouds were fully separated, $\bar{t}_{\rm sep}$, 
were then determined; and then the expectation value of $\bar{x}$ was 
computed numerically for each cloud separately at each time step in the 
range $\bar{t}\ge\bar {t}_{\rm sep}$:
\begin{eqnarray}
\bar{x}_{\rm slow}(\bar{t}) 
&=& 
\int_{-\bar{L}/2}^{\bar{x}_{\rm mid}}
\bar{x}
\left|\bar{\Psi}(\bar{x},\bar{t})\right|^{2}d\bar{x}\nonumber\\
&\equiv&
\bar{x}_{\rm slow}(\bar{t}_{\rm sep}) + 
\bar{v}_{\rm slow}(\bar{t} - \bar{t}_{\rm sep})\nonumber\\
\bar{x}_{\rm fast}(\bar{t})
&=& 
\int^{\bar{L}/2}_{\bar{x}_{\rm mid}}
\bar{x}
\left|\bar{\Psi}(\bar{x},\bar{t})\right|^{2}d\bar{x}\nonumber\\
&\equiv&
\bar{x}_{\rm fast}(\bar{t}_{\rm sep}) + 
\bar{v}_{\rm fast}(\bar{t} - \bar{t}_{\rm sep})
\end{eqnarray}
where $\bar{L}$ is the length of the numerical grid used in the GP
simulation.  Care was taken to make sure that none of the clouds got close
to the edges of the grid.  Finally, the velocities of the fast and slow
clouds were extracted by fitting straight lines to the $\bar{x}_{\rm slow}$
and $\bar{x}_{\rm fast}$ results to obtain $\bar{v}_{\rm slow}$ and $\bar{v}_
{\rm fast}$. The extra velocity "kicks" for each cloud due to repulsion 
were determined by $\delta\bar{v}_{\rm fast}=\bar{v}_{\rm fast}-\bar{v}_{L}$ 
and $\delta\bar{v}_{\rm slow}=\bar{v}_{\rm slow}$.  Convergence runs of 
the GP solver for finer space and time steps showed that $\delta\bar{v}_
{\rm slow}=\delta\bar{v}_{\rm fast}\equiv\delta\bar{v}$ over the entire
interaction strength range.  This shows that momentum was conserved and 
supports the picture of two equal--mass clouds pushing against each other 
as they separate. 

\begin{figure}[t]
\begin{center}
\mbox{\psboxto(3.4in;0.0in){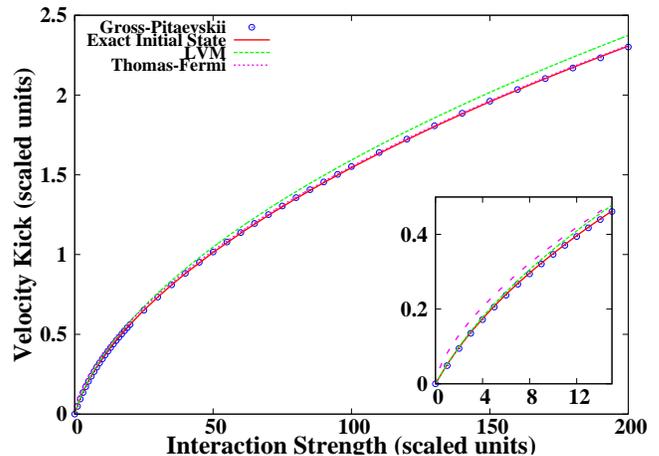}}
\end{center}
\caption{(color online) This figure shows the velocity ``kick'', 
$\delta\bar{v}$, versus interaction strength, $\bar{g}N$.  The open 
circles are the results obtained from GP--equation simulations; the 
three curves show estimates of the velocity kick obtained by equating 
the difference in total kinetic energy of the interacting system
and the non--interacting system with 2/3 of the total interaction
energy.  These estimates are derived from (1) the exact GP initial state
(solid curve), (2) LVM--approximate initial state (dashed curve) and
(3) Thomas--Fermi approximate initial state (dotted curve).}
\label{kick_fig}
\end{figure}

Fig.\ \ref{kick_fig} contains a graph of the velocity kick versus interac%
tion strength for $0\le\bar{g}N\le 200$.  For this set of simulations, the
Bragg pulse was applied and the trap was turned off simultaneously at $\bar{t}
=0$.  For the results shown, the recoil velocity was $\bar{v}_{L}=10$ scaled
velocity units.  In the example mentioned above this would give $^{87}$Rb 
atoms a recoil velocity of 0.15 cm/s.  The result displayed in Fig.\ \ref{%
kick_fig} can be quantitatively understood in a simple way as we now explain.

The velocity kick, $\delta\bar{v}$, is the result of mutual repulsion 
between the fast and slow clouds as they separate after the Bragg pulse.  
From a classical viewpoint, this repulsion will result in a change in the
total kinetic energy of the center--of--mass (CM) motion of the two clouds.  
Thus the difference between the total CM kinetic energy of the two inter%
acting clouds and the kinetic energy of the two non--interacting clouds
should be equal to the energy available for repulsion assuming the repulsion
produces no distortion.

Since the Bragg pulse splits the $N$--atom condensate into two equal pieces 
we can express this equality as (we begin in SI units and then convert):
\begin{eqnarray}
U_{\rm rep}
&=&
\left(\tfrac{N}{2}\right)
\tfrac{1}{2}m\left(v_{L}+\delta v\right)^{2} +
\left(\tfrac{N}{2}\right)
\tfrac{1}{2}m(\delta v)^{2}\nonumber\\
&-&
\left(\tfrac{N}{2}\right)
\tfrac{1}{2}mv_{L}^{2}
\label{ke_diff}
\end{eqnarray}
where $U_{\rm rep}$ is the many--body energy available for different clouds 
to repel each other.  We can derive an expression for this from the total 
many--body interaction energy given by~\cite{Pethick_and_Smith}
\begin{equation}
U_{\rm int} =
\tfrac{1}{2}N(N-1)g\int_{-\infty}^{\infty}dx\,
\left|\Psi(x,t)\right|^{4}
\label{u_int_mb}
\end{equation}
where $\Psi(x,t)$ is the condensate wave function after the $\pi/2$ 
Bragg pulse and has the general form
\begin{equation}
\Psi(x,t) = 
\psi_{1}(x,t) + 
e^{i\Delta kx}\psi_{2}(x,t).
\label{bragg_pi_form}
\end{equation}
We have chosen to distinguish between $\psi_{1}$ and $\psi_{2}$, even though
they are the same in our model, for bookkeeping purposes that will become
apparent below.  Substituting Eq.\ (\ref{bragg_pi_form}) into 
(\ref{u_int_mb}) we have
\begin{eqnarray}
U_{\rm int}
&=&
\tfrac{g}{2}N(N-1)\int_{-\infty}^{\infty}dx
\bigg[
 |\psi_{1}|^{4} + 
 |\psi_{2}|^{4} + 
4|\psi_{1}|^{2}|\psi_{2}|^{2}
\nonumber\\
&+&
\bigg\{
2\left(|\psi_{1}|^{2}+|\psi_{2}|^{2}\right)
\psi_{1}^{\ast}\psi_{2}e^{i\Delta kx} +
\left(\psi_{1}^{\ast}\psi_{2}\right)^{2}e^{2i\Delta kx}\nonumber\\
&+&
{\rm c.c.}
\bigg\}
\bigg]
\label{U_int_all}
\end{eqnarray}
Here we will make a crucial approximation.  This approximation will be 
tested by comparison of the velocity kicks predicted here with those 
determined by numerical solution of the GP equation and will be used 
again in deriving a two--cloud LVM model.

We assume here that $\Delta k$ is large enough so that all of the 
integrals containing exponentials such as $\exp(\pm i\Delta kx)$ and 
$\exp(\pm 2i\Delta kx)$ in Eq.\ (\ref{U_int_all}) can be neglected. This
is equivalent to assuming that the wavelength of the Bragg pulse light
is small compared to the size of the condensate.  Hence terms having
these exponentials oscillate rapidly and the integrals containing them
approximately average to zero.  The result of this approximation is that
all of the terms inside the curly braces in Eq.\ (\ref{U_int_all}) can
be neglected and we can write
\begin{eqnarray}
U_{\rm int}
&\approx&
\tfrac{g}{2}N(N-1)\int_{-\infty}^{\infty}dx
\left(
 |\psi_{1}|^{4} + 
 |\psi_{2}|^{4} + 
4|\psi_{1}|^{2}|\psi_{2}|^{2}
\right)\nonumber\\
&\equiv&
U_{\rm self,1} + U_{\rm self,2} + U_{\rm rep}.
\label{U_int_approx}
\end{eqnarray}

This last expression suggests a picture of the evolution of fast
and slow clouds during separation.  This picture depends on two assumptions:
(1) the separating clouds do not distort significantly so that the form of 
the wave function in Eq.\ (\ref{bragg_pi_form}) is maintained and, (2) the
wavevector, $\Delta k$, is large enough so that the approximation in
Eq.\ (\ref{U_int_approx}) is valid.  In this case, the fast/slow cloud
evolution during separation divides into three categories: (1) self 
interaction of the fast cloud, (2) self interaction of the slow cloud,
and (3) fast/slow cloud interaction.  The energy available for this last
interaction is suggested by the above equation:
\begin{eqnarray}
U_{\rm rep} 
&=& 
\tfrac{1}{2}gN(N-1)
\int_{-\infty}^{\infty}4|\psi_{1}|^{2}|\psi_{2}|^{2}dx\nonumber\\
&\approx&
\tfrac{1}{2}gN^{2}
\int_{-\infty}^{\infty}|\psi|^{4}dx =
\tfrac{2}{3}U_{\rm int}.
\end{eqnarray}
Here we assume that $N\gg 1$ and that $\psi_{1}\approx\psi_{2}\approx\psi/
\sqrt{2}$ where $\psi$ is the (unit norm) condensate wave function {\em 
before} application of the Bragg pulse as in Eqs.\ (\ref{slow_clouds})
and (\ref{fast_clouds}).

This fact also leads to the last equality because, from Eq.\,(\ref{U_int%
_approx}), it is clear that the total energy of interaction is one part 
self interaction of cloud 1, one part self interaction of cloud 2, and 
4 parts cloud--cloud interaction.  The self interaction terms, according 
to our picture, are energies available for expansion while the cloud--cloud
interaction either distorts the cloud shapes and/or changes the velocities 
of their centers--of--mass.  We assume no distortion so all of this energy 
is assumed available for giving the clouds a velocity kick.  We can now
derive this kick by substituting the approximate expression for $U_{\rm rep}$
into Eq.\ (\ref{ke_diff}) and canceling common factors of $N/2$:
\begin{eqnarray}
gN\int_{-\infty}^{\infty}|\psi|^{4}dx
&=&
\tfrac{1}{2}m\left(v_{L}+\delta v\right)^{2} +
\tfrac{1}{2}m(\delta v)^{2}\nonumber\\
&-&
\tfrac{1}{2}mv_{L}^{2}\nonumber\\
\bar{g}N\int_{-\infty}^{\infty}|\bar{\psi}|^{4}d\bar{x}
&=&
\tfrac{1}{4}\left(\bar{v}_{L}+\delta\bar{v}\right)^{2} +
\tfrac{1}{4}(\delta\bar{v})^{2}\nonumber\\
&-&
\tfrac{1}{4}\bar{v}_{L}^{2},
\label{kick_eq}
\end{eqnarray}
where in second line above we have converted back to scaled units.  Thus
we can now write an expression for the velocity kick:
\begin{equation}
\delta\bar{v} =
\left[
\tfrac{1}{4}\bar{v}_{L}^{2} + 2\bar{u}_{sp}
\right]^{1/2} - \tfrac{1}{2}\bar{v}_{L},
\label{dv_formula}
\end{equation}
where
\begin{equation}
\bar{u}_{sp} \equiv
\bar{g}N\int_{-\infty}^{\infty}|\bar{\psi}|^{4}d\bar{x}
\label{u_sp}
\end{equation}
and $\bar{v}_{L}=2\Delta\bar{k}$.  We can think of the separating 
clouds being pushed apart by a spring in between them and $\bar{u}_{sp}$ 
is the initial energy stored in the spring.

The comparison of values of $\delta\bar{v}$ predicted by the above model
with those determined from numerical solution of the GP equation are
shown in Fig.\ \ref{kick_fig}.  The discrete points are the numerically
determined values and the remaining three curves are computed via Eq.\ %
(\ref{dv_formula}) where $\bar{u}_{sp}$ has been calculated using three 
expressions for $\psi$, the initial state condensate wave function.  These
three versions of $\psi$ were (1) the exact initial state from the GP
simulation (solid curve), (2) an LVM gaussian cloud where the gaussian width 
was the stationary value for a 1D condensate confined in the initial 
harmonic trap (dashed line), and (3) the Thomas--Fermi--approximate solution 
of the GP equation for the trapped condensate (dotted line).   All three
expressions, as can be seen from the graph, gave good estimates. We found 
that using the exact GP initial state gave near--perfect agreement (to 
four decimal places) with the kick determined from numerical GP.  The LVM
gaussian performed very well for small $\bar{g}N$ and less well for large 
while the Thomas--Fermi works well for large $\bar{g}N$ and less well for 
small.

This agreement between the numerically determined kicks and the estimates
from our heuristic model lends support to our picture of the effect of 
cloud/cloud interaction during separation.  We will use this in what follows
to develop a two--cloud LVM technique for modeling a full Bragg 
interferometer.  
 
\section{Two--cloud LVM model}
\label{two_cloud}

In this section we present a two--cloud LVM model to approximate the
evolution of the condensate wave function following a $\pi/2$ Bragg
pulse.  The presence of two clouds will enable the model to account for
cloud--cloud interactions and should be able to predict the velocity
kick accurately.  Below we present both 1D and 3D versions of this model.
This will enable us to make quantitative comparisons of our model with
the results of 1D GP simulations of the entire $\pi/2$--$\pi$--$\pi/2$
interferometer.  The 3D model will be useful for assessment of real--world
AI designs.  In addition, we also present comparisons of our 3D two--cloud
model with the results of the original NIST experiment.

\subsection{1D two--cloud LVM}
\label{1D_two_cloud}

The trial wave function for the 1D two--cloud LVM model is taken to be
a sum of two 1D gaussian wavepackets where one of them is boosted to
a velocity of $\bar{v}_{L}=2\Delta\bar{k}$:
\begin{equation}
\bar{\Psi}(\bar{x},\bar{t}) = 
\frac{\bar{A}_{1D}(\bar{t})}{\sqrt{2}}
\left(
e^{f_{1}\left(\bar{x},\bar{t}\right)} +
e^{i\Delta\bar{k}\bar{x}}
e^{f_{2}\left(\bar{x},\bar{t}\right)}
\right)
\label{two_cloud_trial_wf}
\end{equation}
where
\begin{equation}
f_{j}(\bar{x},\bar{t}) 
\equiv
-\frac{\left(\bar{x}-\bar{x}_{j}(\bar{t})\right)^{2}}
{2\bar{w}^{2}(\bar{t})} +
i\bar{\alpha}_{j}(\bar{t})x + i\bar{\beta}(\bar{t})x^{2}
\end{equation}
where $j = 1,2$.  We note here that both the slow cloud (cloud 1) and 
the fast cloud share the same width, $\bar{w}$, and quadratic phase 
curvature, $\bar{\beta}$ but have differing centers and linear phase
coefficients.  This assumption is borne out in multiple GP simulations
as can be seen, for example, in Fig.\ \ref{bp_evol}.  Imposing the 
normalization condition yields the following constraint: $|\bar{A}_{1D}|^{2}
\bar{w}\pi^{1/2}=1$ where all terms containing exponentials such as 
$e^{\pm i\Delta\bar{k}\bar{x}}$ were neglected.

Carrying out the procedure described in Section \ref{models_overview},
we calculate the LVM Lagrangian by inserting the trial wave function
into the Lagrangian density and integrating. If we neglect rapidly 
oscillating terms and impose the normalization constraint we obtain the
following result.
\begin{eqnarray}
\bar{L}_{1D} &=&
\tfrac{1}{2}\dot{\bar{\alpha}}_{1}\bar{x}_{1} +
\tfrac{1}{2}\dot{\bar{\alpha}}_{2}\bar{x}_{2} +
\tfrac{1}{2}\dot{\bar{\beta}}
\left(\bar{x}_{1}^{2} + \bar{x}_{2}^{2} + \bar{w}^{2}\right)\nonumber\\
&+&
\tfrac{1}{2\bar{w}^{2}} +
\tfrac{1}{2}\left(\bar{\alpha}_{1} + 
2\bar{\beta}\bar{x}_{1}\right)^{2} +
\tfrac{1}{2}\left(\bar{\alpha}_{2} + 
2\bar{\beta}\bar{x}_{2}\right)^{2}\nonumber\\
&+&
(\Delta\bar{k})\left(\bar{\alpha}_{2}+2\bar{\beta}\bar{x}_{2}\right) +
\tfrac{1}{2}\left(\Delta\bar{k}\right)^{2} + 
2\bar{\beta}^{2}\bar{w}^{2}\nonumber\\
&+&
\tfrac{1}{8}\gamma^{2}
\left(\bar{x}_{1}^{2}+\bar{x}_{2}^{2}+\bar{w}^{2}\right)\nonumber\\ 
&+&
\left(\tfrac{\bar{g}N}{4\left(2\pi\right)^{1/2}\bar{w}}\right)
\left(
1 + 2\,e^{-\left(\bar{x}_{1}-\bar{x}_{2}\right)^{2}/2\bar{w}^{2}}
\right),
\end{eqnarray}
where $\gamma=\omega/\omega_{0}$ is the ratio of the actual trap frequency
to the frequency used to define the length unit.  

We find the equations of motion using the ordinary Euler--Lagrange equations
(see Eq.\ (\ref{E_L_eqs})).  After some rearrangement these equations can 
be expressed as follows:
\begin{eqnarray}
\ddot{\bar{x}}_{1} + \gamma^{2}\bar{x}_{1}
&=&
\ \ F_{12}^{(1D)}(\bar{x}_{1},\bar{x}_{2},\bar{w})\nonumber\\
\ddot{\bar{x}}_{2} + \gamma^{2}\bar{x}_{2}
&=&
-F_{12}^{(1D)}(\bar{x}_{1},\bar{x}_{2},\bar{w})\nonumber\\
\ddot{\bar{w}} + \gamma^{2}\bar{w}
&=& 
\ \ F_{w}^{(1D)}(\bar{x}_{1},\bar{x}_{2},\bar{w})
\label{1d_2cloud_xw}
\end{eqnarray}
where
\begin{equation}
F_{12}^{(1D)}(\bar{x}_{1},\bar{x}_{2},\bar{w}) \equiv
\left(\tfrac{2\bar{g}N}{\left(2\pi\right)^{1/2}\bar{w}^{3}}\right)
\left(\bar{x}_{1}-\bar{x}_{2}\right)
e^{-\left(\bar{x}_{1}-\bar{x}_{2}\right)^{2}/2\bar{w}^{2}}
\label{1d_2cloud_xforce}
\end{equation}
and
\begin{eqnarray}
F_{w}^{(1D)}(\bar{x}_{1},\bar{x}_{2},\bar{w})
&\equiv&
\tfrac{4}{\bar{w}^{3}} +
\left(\tfrac{\bar{g}N}{\left(2\pi\right)^{1/2}\bar{w}^{2}}\right)\nonumber\\
&\times&
\left[
1 + 2
\left(1-\tfrac{\left(\bar{x}_{1}-\bar{x}_{2}\right)^{2}}{\bar{w}^{2}}\right)
e^{-\left(\bar{x}_{1}-\bar{x}_{2}\right)^{2}/2\bar{w}^{2}}
\right]\nonumber\\
\label{1d_2cloud_wforce}
\end{eqnarray}
where $F_{12}^{(1D)}$ can be thought of as the ``force'' of repulsion between
the separating clouds and $F_{w}^{(1D)}$ roughly thought of as the ``force''
causing the width to change. 

The above equations constitute a closed system of equations to be solved
for $\bar{x}_{1}$, $\dot{\bar{x}}_{1}$, $\bar{x}_{2}$ , $\dot{\bar{x}}_{1}$,
$\bar{w}$, and $\dot{\bar{w}}$.  Once these equations are solved, the 
remaining parameters appearing in the trial wave function can be obtained
as follows.
\begin{eqnarray}
\bar{\beta} &=& \tfrac{\dot{\bar{w}}}{4\bar{w}}\nonumber\\
\bar{\alpha}_{1} &=& 
\tfrac{1}{2}\dot{\bar{x}}_{1} - 2\bar{\beta}\bar{x}_{1}\nonumber\\
\bar{\alpha}_{2} &=& 
\tfrac{1}{2}\dot{\bar{x}}_{2} - 2\bar{\beta}\bar{x}_{2} - \Delta\bar{k}
\label{1d_2cloud_ab}
\end{eqnarray}

The above system of equations have some interesting properties that are
analogous to those for Newton's second law.  For example, if we introduce
the center of mass of the two--cloud system, $\bar{x}_{cm}\equiv (\bar{x}_{1}
+\bar{x}_{2})/2$, and the relative coordinate, $\bar{x}_{rel}\equiv \bar{x}_{1}
-\bar{x}_{2}$ we find their equations of motion by adding and subtracting
the first two of Eqs.\ (\ref{1d_2cloud_xw}) respectively.  They are
\begin{eqnarray}
\ddot{\bar{x}}_{cm} + \gamma^{2}\bar{x}_{cm} &=& 0\nonumber\\
\ddot{\bar{x}}_{rel} + \gamma^{2}\bar{x}_{rel} &=& 
2F_{12}^{(1D)}(\bar{x}_{1},\bar{x}_{2},\bar{w}) \equiv 2F_{12}^{(1D)}(\bar{x}_{rel},\bar{w})\nonumber\\
\label{1d_2cloud_cmrel}
\end{eqnarray} 
where, in the last equality, we noted that $F_{12}^{(1D)}$ only depends on the
relative coordinate and the width.  The first of the above equations can 
be solved immediately by inspection and is equivalent to the classical 
result that the CM motion of a system only depends on external forces.
This reduces the total number of equations that must be solved numerically
to the equation for $\bar{w}$ in Eqs.\ (\ref{1d_2cloud_xw}) and the equation
for $\bar{x}_{rel}$ above.

There is also a conserved ``energy'' that can be written as
\begin{equation}
\bar{E}_{1D} =
\tfrac{1}{2}\dot{\bar{x}}_{1}^{2} +
\tfrac{1}{2}\dot{\bar{x}}_{2}^{2} +
\tfrac{1}{2}\dot{\bar{w}}^{2} +
\bar{U}_{1D}(\bar{x}_{1},\bar{x}_{2},\bar{w})
\label{E_1D}
\end{equation}
where the ``potential energy'' $\bar{U}_{1D}$ is given by
\begin{eqnarray}
\bar{U}_{1D}(\bar{x}_{1},\bar{x}_{2},\bar{w}) 
&=& 
\tfrac{1}{2}\gamma^{2}\bar{x}_{1}^{2} +
\tfrac{1}{2}\gamma^{2}\bar{x}_{2}^{2} +
\tfrac{1}{2}\gamma^{2}\bar{w}^{2} +
\tfrac{2}{\bar{w}^{2}}\nonumber\\ 
&+&
\left(\tfrac{\bar{g}N}
{\left(2\pi\right)^{1/2}\bar{w}}\right)
\left(
1 + 2\,e^{-(\bar{x}_{1}-\bar{x}_{2})^{2}/2\bar{w}^{2}}
\right).\nonumber\\
\label{U_1D}
\end{eqnarray}
Equations (\ref{1d_2cloud_xw}) can all be written as the second time
derivative of each coordinate equals the negative partial derivative of 
the above potential energy with respect to the corresponding coordinate
(e.g., $\ddot{\bar{x}}_{1}=-\partial\bar{U}_{1D}/\partial\bar{x}_{1}$).  
This constant of the motion can be used to estimate the velocity kick 
and is also useful as a check on numerics.  

\begin{figure}[t]
\begin{center}
\mbox{\psboxto(3.5in;0.0in){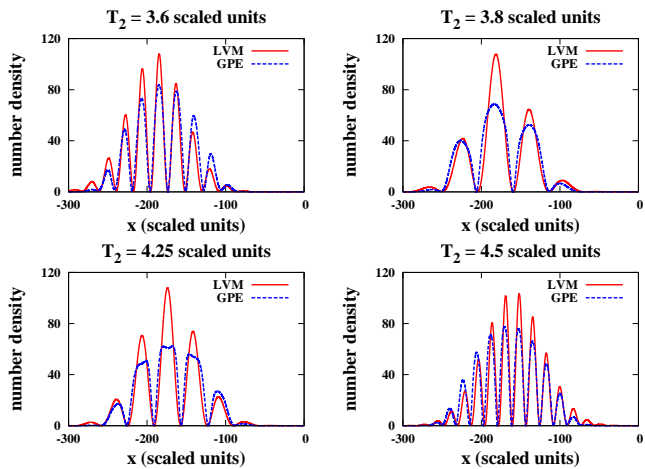}}
\end{center}
\caption{(color online) These plots show a comparison of the interference
patterns resulting from a $\pi/2$--$\pi$--$\pi/2$ Bragg AI computed by the 
GP equation (dashed line) and by the Bragg prototyper method.  The
interferometer times were (in scaled time units) $T_{0}=1$, $T_{1}=4$, 
$T_{2}$ variable, and $T_{3}=13.4$.  The times for $T_{2}$ are (clockwise
from upper left) $T_{2}=3.6,3.8,4.25,4.5$ scaled time units.  The 
interaction strength for all runs was $\bar{g}N=10$ scaled units.}
\label{1d_gp_cmp}
\end{figure} 

Equations (\ref{1d_2cloud_xw}) and (\ref{1d_2cloud_ab}) can be used with 
the rest of the Bragg prototyper model to simulate Bragg AI behavior. We 
have used this 1D Bragg prototyper model to predict the results of a 
$\pi/2$--$\pi$--$\pi/2$ Bragg AI and have also simulated such an AI with
the 1D GP equation.  In these runs, the condensate was released from the
trap and allowed to expand for $T_{0}=1$ scaled time unit.  The $\pi/2$
Bragg pulse was then applied and, after a time interval $T_{2}=4$ scaled
units, a $\pi$ pulse was applied followed by a variable time interval,
$T_{2}$, at which time the final Bragg pulse was turned on.  The clouds
were allowed to evolve for an additional $T_{3}=13.4$ time units.  Figure
\ref{1d_gp_cmp} shows the comparison of the interference patterns predicted
by the Bragg prototyper model and by the GP equation.  Each graph 
corresponds to a different value of $T_{2}$.  In the figure, the values
were (clockwise from upper left panel) $T_{2}=3.6,3.8,4.25,4.5$ time units.

\begin{figure*}[t]
\begin{center}
\mbox{\psboxto(\textwidth;0.0in){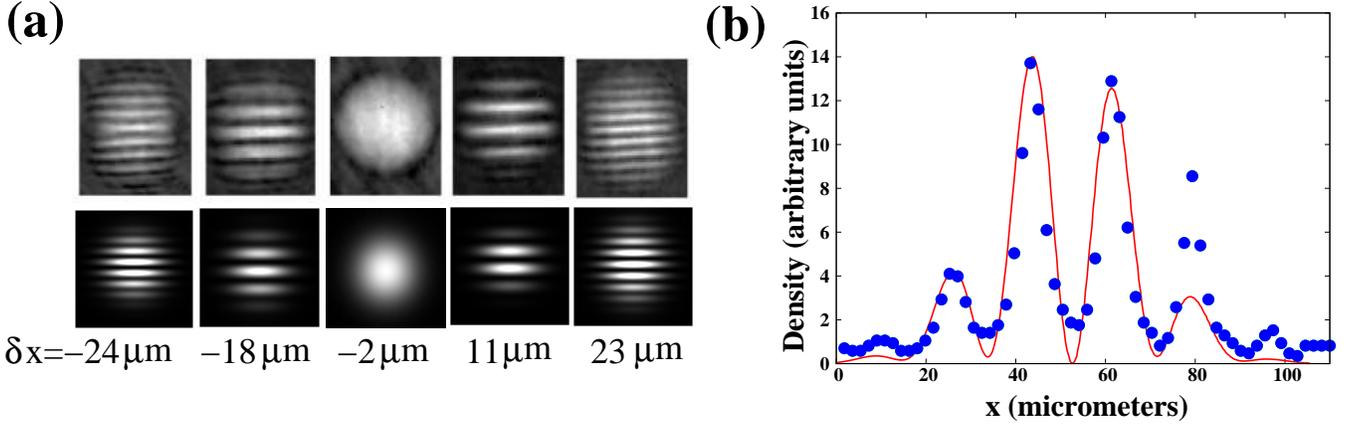}}
\end{center}
\caption{(color online) This figure exhibits a comparison of the results
of the 3D two--cloud Bragg AI model with the experimental results given in
Fig.\ 2 of Ref.\ \cite{PhysRevLett.85.2040} where the NIST experiment 
discussed earlier in the text was described.  In that experiment, the 
evolving phase of a Bose--Einstein condensate wave function was probed using
a $\pi/2$--$\pi$--$\pi/2$ Bragg interferometer. (a) Interference patterns 
obtained for interferometer runs (top row experiment, bottom row theory) 
where $T_{0}=4$ ms, $T_{1}=1$ ms, $T_{3}=2$ ms, and $T_{2}$ was varied so 
that the cloud--center spacings ($\delta x$) at the last Bragg pulse were
the values given in the figure; (b) A quantitative comparison of the case 
for $\delta x=11\mu$m (second from right end in panel (a)).  The theory 
curve was normalized to match the highest experimental peak.}
\label{3d_exp_cmp}
\end{figure*} 

The GP equation was solved using a Crank--Nicolson algorithm on a space 
grid of width $\bar{L}=800$ length units and divided into $N_{x}=32768$
space steps over a total time of $\bar{t}_{max}=22$ scaled time units using
$N_{t}=1200000$ time steps.  The initial state was computed by integrating 
the GP equation in imaginary time with the trap on.  The time needed to 
obtain fully converged final interference patterns was about five hours of 
run time on a commodity laptop.  The time required to obtain the model 
results on the same computer was about 1 second, approximately 18,000 times
faster.  We expect that the speedup factor for 3D solutions to be 2--3
orders--of--magnitude greater.  This makes our model an essential tool for
assessment of more complicated atom interferometers which result from new 
AI designs inspired by quantum information science.

It is easy to see that the Bragg prototyper model reproduces the number
and spacing of the GP--equation interference fringes in all cases.  The
major difference is that the width and height of the GP fringes are wider 
and shorter, respectively, than those in the LVM pattern.  It is likely that
the area under the curve in corresponding GP and LVM fringes is the same.
This would make wider fringes lower and would imply that the number of
atoms in each GP fringe was approximately equal to the number of atoms
in the corresponding LVM fringe. 

\subsection{3D two--cloud LVM}
\label{3D_two_cloud}

We also derived a 3D version of the two--cloud LVM model.  The derivation
is a straightforward generalization of the 1D version with all of the same
assumptions and approximations.  We include the highlights of its derivation
for completeness.

The trial wavefunction for the 3D two--cloud LVM model is a sum of two
3D gaussians:
\begin{equation}
\Psi(\bar{\bf r},\bar{t}) = 
\frac{\bar{A}_{3D}}{\sqrt{2}}
\left(
e^{F_{1}\left(\bar{\bf r},\bar{t}\right)} +
e^{i\Delta\bar{\bf k}\cdot\bar{\bf r}}
e^{F_{2}\left(\bar{\bf r},\bar{t}\right)}
\right)
\label{3d_2cloud_trial_wf}
\end{equation}
where
\begin{equation}
F_{j}(\bar{\bf r},\bar{t}) = 
\sum_{\eta=x,y,z}
\left(
-\frac{(\bar{\eta}-\bar{\eta}_{j})^{2}}{2\bar{w}_{\eta}^{2}} +
i\left(\bar{\alpha}_{j\eta}\bar{\eta} +
\bar{\beta}_{\eta}\bar{\eta}^{2}\right)
\right)
\end{equation}
and $j=1,2$.  In three dimensions we allow for the possibility of differing
widths, $\bar{w}_{\eta},\ \eta=x,y,z$ along the three axes, however we 
assume that these widths are the same for both clouds.  This holds for the
corresponding phase curvature coefficients $\bar{\beta}_{\eta},\ \eta=x,y,z$
as well.  It is straightforward to calculate the normalization constraint
as $|\bar{A}_{3D}|^{2}\bar{w}_{x}\bar{w}_{y}\bar{w}_{y}\pi^{1/2}=1$.  We 
note again that it was necessary to neglect rapidly oscillating terms that
contained exponentials such as $e^{\pm i\Delta\bar{k}\cdot\bar{\bf r}}$
to arrive at this result.

\begin{figure*}[t]
\begin{center}
\mbox{\psboxto(\textwidth;0.0in){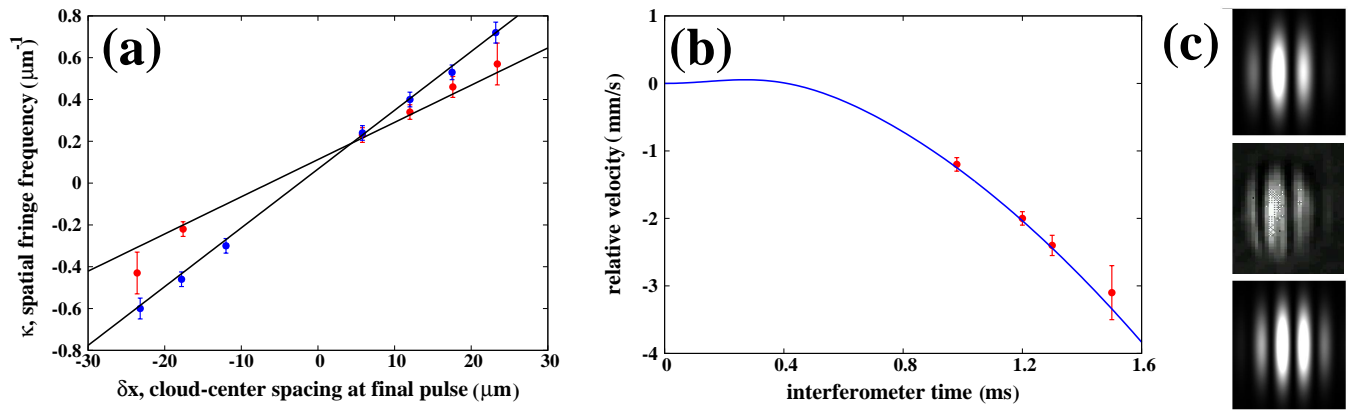}}
\end{center}
\caption{(color online) (a) The spatial fringe frequency versus the cloud separation at the instant the final $\pi/2$ pulse is applied. The data is 
taken from Fig.\ 3 of Ref.\ \cite{PhysRevLett.85.2040}.  The steeper line corresponds to $T_{0}=4$ ms, and the other line to $T_{0}=1$ ms. (b) The 
relative velocity between interfering clouds versus $T$, the interferometer
time. Data is taken from Fig.\ 5 of the cited paper. (c) An interference 
pattern comparison between 3D one-cloud LVM (bottom), experiment (middle), 
and 3D two-cloud LVM (top), with the trap on during all the steps of the
interferometer and $T=1$ ms.}
\label{3d_exp_cmp_more}
\end{figure*} 

The Lagrangian for this trial wave function thus becomes
\begin{eqnarray}
\bar{L}_{3D} &=&
\sum_{\eta=x,y,z}
\bigg\{
\tfrac{1}{2}\dot{\bar{\alpha}}_{1\eta}\bar{\eta}_{1} +
\tfrac{1}{2}\dot{\bar{\alpha}}_{2\eta}\bar{\eta}_{2} +
\tfrac{1}{2}\dot{\bar{\beta}}_{\eta}
\left(
\bar{\eta}_{1}^{2}+
\bar{\eta}_{2}^{2}+
\bar{w}_{\eta}^{2}
\right)\nonumber\\
&+&
\tfrac{1}{2}\left(\bar{\alpha}_{1\eta}+
2\bar{\beta}_{\eta}\bar{\eta}_{1}\right)^{2} +
\tfrac{1}{2}\left(\bar{\alpha}_{2\eta}+
2\bar{\beta}_{\eta}\bar{\eta}_{2}+
\Delta\bar{k}_{\eta}\right)^{2}\nonumber\\
&+&
\tfrac{1}{2\bar{w}_{\eta}^{2}} +
2\bar{\beta}_{\eta}^{2}\bar{w}_{\eta}^{2} +
\tfrac{1}{8}\gamma_{\eta}^{2}
\left(
\bar{\eta}_{1}^{2}+
\bar{\eta}_{2}^{2}+
\bar{w}_{\eta}^{2}
\right)
\bigg\}\nonumber\\
&+&
\left(
\tfrac{\bar{g}N}{4\left(2\pi\right)^{3/2}
\bar{w}_{x}\bar{w}_{y}\bar{w}_{z}}
\right)
\bigg(1 + 
2\,e^{-\sum_{\eta=x,y,z}
\tfrac{(\bar{\eta}_{1}-\bar{\eta}_{2})^{2}}
{2\bar{w}_{\eta}^{2}}}
\bigg)\nonumber\\
\end{eqnarray}
where $\gamma_{\eta}\equiv\omega_{\eta}/\bar{\omega}$ and $\bar{\omega}$
is the geometric average of the three trap frequencies that is used in the
definition of the 3D length unit.

The equations of motion that arise from the above Lagrangian are
\begin{eqnarray}
\ddot{\bar{\eta}}_{1}+\gamma_{\eta}^{2}\bar{\eta}_{1}
&=& 
\ \ F_{12\eta}^{(3D)}
\left(\bar{\bf r}_{1},\bar{\bf r}_{2},\bar{\bf w}\right)\nonumber\\
\ddot{\bar{\eta}}_{2}+\gamma_{\eta}^{2}\bar{\eta}_{2}
&=& 
-F_{12\eta}^{(3D)}
\left(\bar{\bf r}_{1},\bar{\bf r}_{2},\bar{\bf w}\right)\nonumber\\
\ddot{\bar{w}}_{\eta}+\gamma_{\eta}^{2}\bar{w}_{\eta}
&=& 
\ \ F_{w\eta}^{(3D)}
\left(\bar{\bf r}_{1},\bar{\bf r}_{2},\bar{\bf w}\right)
\quad
\eta=x,y,z.
\label{3d_2cloud_xw}
\end{eqnarray}
where $\bar{\bf r}_{j}\equiv\left(\bar{x}_{j},\bar{y}_{j},\bar{z}_{j}\right)$
with $j=1,2$ and 
$\bar{\bf w}\equiv\left(\bar{w}_{x},\bar{w}_{y},\bar{w}_{z}\right)$ and
where the ``force'' terms on the right--hand--sides above are given by
\begin{eqnarray}
F_{12\eta}^{(3D)}\left(\bar{\bf r}_{1},\bar{\bf r}_{2},\bar{\bf w}\right)
&=&
\left(
\tfrac{2\bar{g}N}{\left(2\pi\right)^{3/2}
\bar{w}_{x}\bar{w}_{y}\bar{w}_{z}\bar{w}_{\eta}}
\right)
\left(
\tfrac{\bar{\eta}_{1}-\bar{\eta}_{2}}
{\bar{w}_{\eta}}
\right)\nonumber\\
&\times&
e^{-\sum_{\eta=x,y,z}
\tfrac{(\bar{\eta}_{1}-\bar{\eta}_{2})^{2}}
{2\bar{w}_{\eta}^{2}}}\nonumber\\
F_{w\eta}^{(3D)}\left(\bar{\bf r}_{1},\bar{\bf r}_{2},\bar{\bf w}\right)
&=&
\tfrac{4}{\bar{w}_{\eta}^{3}} +
\left(
\tfrac{\bar{g}N}{\left(2\pi\right)^{3/2}
\bar{w}_{x}\bar{w}_{y}\bar{w}_{z}\bar{w}_{\eta}}
\right)
\bigg[1 +\nonumber\\ 
&2&
\times
\left(
1-\tfrac{\left(\bar{\eta}_{1}-\bar{\eta}_{2}\right)^{2}}{\bar{w}_{\eta}^{2}}
\right)
e^{-\sum_{\eta=x,y,z}
\tfrac{(\bar{\eta}_{1}-\bar{\eta}_{2})^{2}}
{2\bar{w}_{\eta}^{2}}}
\bigg]\nonumber\\
\label{3d_2cloud_xw_forces}
\end{eqnarray}
The rest of the equations connect the widths, and CM positions and 
velocities to the parameters that actually appear in the trial wave function.
These are

\begin{eqnarray}
\bar{\beta}_{\eta} &=& \tfrac{\dot{\bar{w}}_{\eta}}
{4\bar{w}_{\eta}}\nonumber\\
\bar{\alpha}_{1\eta} &=& 
\tfrac{1}{2}\dot{\bar{\eta}}_{1} - 
2\bar{\beta}_{\eta}\bar{\eta}_{1}\nonumber\\
\bar{\alpha}_{2\eta} &=& 
\tfrac{1}{2}\dot{\bar{\eta}}_{2} - 
2\bar{\beta}_{\eta}\bar{\eta}_{2} - \Delta\bar{k}_{\eta}\nonumber\\
\eta &=& x,y,z.
\label{3d_2cloud_ab}
\end{eqnarray}

The 3D version of the 2--cloud model shares some of the properties 
associated with the 1D version.  These include (1) the motion of the CM,
$\bar{\bf r}_{cm}=(\bar{\bf r}_{1}+\bar{\bf r}_{2})/2$ is governed only 
by the trapping force and, (2) there is a conserved energy given by
\begin{equation}
\bar{E}_{3D} =
\tfrac{1}{2}\sum_{\eta=x,y,z}
\left\{
\dot{\bar{\eta}}_{1}^{2} +
\dot{\bar{\eta}}_{2}^{2} +
\dot{\bar{w}}_{\eta}^{2}
\right\} +
\bar{U}_{3D}(\bar{\bf r}_{1},\bar{\bf r}_{2},\bar{\bf w})
\label{E_3D}
\end{equation}
where
\begin{eqnarray}
\bar{U}_{3D}(\bar{\bf r}_{1},\bar{\bf r}_{2},\bar{\bf w})
&=& 
\sum_{\eta=x,y,z}
\left\{\tfrac{1}{2}
\gamma_{\eta}^{2}
\left(
\bar{\eta}_{1}^{2} +
\bar{\eta}_{2}^{2} +
\bar{w}_{\eta}^{2}
\right) +
\tfrac{2}{\bar{w}_{\eta}^{2}}
\right\}
\nonumber\\ 
&+&
\left(\tfrac{\bar{g}N}
{\left(2\pi\right)^{3/2}\bar{w}_{x}\bar{w}_{y}\bar{w}_{z}}\right)\nonumber\\
&\times&
\bigg(
1 + 2\,
e^{-\sum_{\eta=x,y,z}
\tfrac{(\bar{\eta}_{1}-\bar{\eta}_{2})^{2}}
{2\bar{w}_{\eta}^{2}}}
\bigg).\nonumber\\
\label{U_3D}
\end{eqnarray}

These equations can now be used as a part of the two--gaussian--cloud
LVM method to predict the behavior of atom interferometers.  We have 
applied this model to the NIST experiment described earlier and detailed
in Ref.\ \cite{PhysRevLett.85.2040}.  Figure \ref{3d_exp_cmp}(a) shows
interference patterns obtained in the experiment (top row) compared with
the model.  The timings of interferometer runs that produced these patterns 
were $T_{0}=4$ ms, $T_{1}=1$ ms, and $T_{3}=2$ ms.  The values of $T_{2}$ 
were varied so that the spacing of the cloud centers, at the moment the 
final Bragg pulse is applied, $\delta x$, was as shown in Fig.\ \ref{%
3d_exp_cmp}(a).  

We see that there is good agreement with the experimental patterns
in terms of number and spacing of fringes.  Figure \ref{3d_exp_cmp}(b)
shows a more quantitative comparison for the case of $\delta x=11\mu$m.
The data was taken from Fig.\ 2 of Ref.\ \cite{PhysRevLett.85.2040} and
the theory is the result of the 3D two--cloud Bragg AI prototyper.  All 
of these experiments were carried out with the trapping potential turned
off.  Since the kick is included naturally in our model, there was no need
to add a velocity kick correction to achieve this level of agreement as 
there was for the one--cloud model.

In the NIST experiment, the spatial fringe frequency, $\kappa$, at the 
last Bragg pulse was measured as a function of $\delta x$ for the trap off 
case.  The comparison of these results with the 3D two-cloud LVM is shown in
Fig.\ \ref{3d_exp_cmp_more}(a) where the data is taken from Fig.\ 3 of
Ref.\ \cite{PhysRevLett.85.2040}.  In our LVM method, the atom density
at the time of the last pulse oscillates as $\cos((\alpha_{2x}-\alpha_{1x}
)x)$.  Thus we have $\kappa_{LVM}=\alpha_{2x}-\alpha_{1x}$ evaluated
at the time of the last pulse.  The above comparison shows excellent 
agreement with experiment.  The relative velocity between the clouds in 
a given pair was also measured for different values of $T_{1}=T_{2}
\equiv T$ when the trap was left on. The data from Fig.\ 5(b) of 
Ref.\ \cite{PhysRevLett.85.2040} is shown in Fig.\ \ref{3d_exp_cmp_more}(b)
along with the results of the LVM for the relative velocity versus $T$. 
Again we find good agreement with the experiment.

The Fig. \ref{3d_exp_cmp_more}(c) shows a comparison of the 3D one-cloud 
LVM (bottom), 3D two-cloud LVM (top), and experiment (middle), for the
interference pattern resulting from an interferometer run in which the 
trap was on and $T=1$ ms. Although the one--cloud and two--cloud LVM inter%
ference patterns are both qualitatively similar to the experimental one,
it is clear that the two--cloud pattern agrees better.  Thus the apparent
agreement between the one--cloud LVM and experiment presented in Ref.\ \cite
{PhysRevLett.85.2040} in the trap--on case without the correction was
fortuitous.

\section{Discussion}
\label{discussion}

In this paper we have presented a method suitable for rapid estimation of
interference patterns deriving from ultra--cold Bragg atom interferometers.
The method achieves this by representing the condensate wave function as a superposition  of fast and slow gaussian clouds and then modeling the changes 
caused by (1) Bragg $\pi/2$ and $\pi$ pulses and (2) by approximating
the GP--equation evolution of the wave function during the intervals between
the pulses.  Thus, by following the sequence of Bragg pulses and intervals 
that occur in a particular Bragg AI, it is possible to approximate rapidly 
the final condensate wave function and thus calculate the expected 
interference pattern.

In this model, Bragg pulses are assumed to change the wave function by
instantaneous shifts in momentum space while {\em rapid} estimation of the
effect of GP evolution is approximated using an LVM technique.  We have
validated the 1D version of this method by comparing its results with 1D
numerical simulations of Mach--Zehnder--type atom interferometers using 
the GP equation.  The 3D version was validated by comparison with experimental
results presented in Ref.\ \cite{PhysRevLett.85.2040}.  We found that the 
method provides good agreement with interference patterns in both the 1D 
and 3D cases as regards the number and spacing of fringes.  In the 1D case,
we found that our model obtained the final interference patterns at least 
10,000 times faster than direct GP--equation simulation on the same computer.
We expect this factor to be several orders--of--magnitude greater for the
3D case.

While the comparison of the results of the method with 1D GP simulations and 
with 3D experiment is quite good, there are some differences.  In the 1D 
case, we found that it was necessary to shift our interference patterns 
over to line up fringe locations with the GP results.  The shifts were
small; they were at most 10 scaled length units which was a small difference
compared to the width of a fringe.  We did not have enough information 
about the experiment to determine if this was necessary in the 3D case.
Also, as can be seen in Fig.\ \ref{1d_gp_cmp}(a), the fringe heights differ 
from the GP pattern.  This difference is also present in the 3D comparison.

There are several possible reasons for these differences.  First, our 
model tacitly assumes that separating clouds are not distorted by their
mutual interaction since we model them as gaussians. In the 1D case, this
assumption holds reasonably but not perfectly well over the range of 
$\bar{g}N$ considered.  GP simulations with larger and larger $\bar{g}N$ 
values show that clouds become more and more distorted as the interaction
strength increases.  Thus some of the energy available for repulsion can be
diverted from changing the CM velocity into cloud distortion.  

Another reason for the difference may lie in our treatment of interacting 
clouds between the final Bragg pulse (which creates two fast and two slow 
clouds) and the time that an image is taken.  During this interval, we use 
the two--cloud model to propagate each fast/slow pair separately and combine
them at the end.  This neglected the interaction of the overlapping clouds 
(the two fast clouds overlap and the two slow clouds overlap) as the two
cloud pairs move apart.  In fact, the $T_{3}$ interval is usually longer
than all of the other intervals combined so there is ample opportunity for
the two clouds in a pair to repel each other.  This could be accounted for
in a four--cloud model.

We also presented a study of the extra velocity kick that the slow and fast
clouds receive when a $\pi/2$ Bragg pulse is applied to a condensate.  When
the condensate splits, the two clouds push against each other as they 
separate.  This causes the fast cloud to acquire an extra velocity, $\delta
\bar{v}$, in addition to the velocity, $\bar{v}_{L}$, imparted by the
light.  The slow cloud recoils at velocity $\delta\bar{v}$ due to conservation
of momentum.  This study was conducted in 1D where the evolution of the
condensate was simulated with the GP equation and the value of $\delta\bar{v}$
was determined directly from the simulation as a function of the interaction
strength $\bar{g}N$.  The values of $\bar{g}N$ ranged from non--interacting
up to well into the Thomas--Fermi regime.  

We found that these velocity kicks could be precisely predicted by setting 
the energy available for repulsion equal to the total kinetic energy of 
the interacting system minus the kinetic energy of the non--interacting 
system.  We approximated the energy available for repulsion from the 
expression for the total interaction energy by neglecting rapidly 
oscillating terms involving the wavevector of the Bragg pulse light.  The
success of this procedure in reproducing the velocity kicks reinforces the
picture of the interaction energy being partitioned into self--interaction 
energy of individual clouds and energy for cloud--cloud interaction.  And
furthermore, that the cloud--cloud interaction only produces a CM velocity
change.

It is also important to mention that there are some phenomena that occur
in Bragg atom interferometers that the model described in this paper might
be modified to handle.  Bragg processes often result in elastic scattering
into intially unoccupied transverse momentum modes~\cite{PhysRevLett.84.5462}.
While these processes are not treated by the GP equation, it is possible
to modify the effect of a pulse in these cases by adding clouds that occupy
these momentum modes and neglecting their interaction with the mother 
condensate during separation.  These modifications may also address other
wave mixing processes that sometimes occur.  Stray light can also cause 
problems for Bragg atom interferometers especially for condensates trapped
near an atom chip.  The major effect of stray light in these cases is to
change the internal energy state of the atom.  The model described above 
can be generalized to account for multiple internal levels of the atoms.

Another difficulty for Bragg interferometers is imperfections in the Bragg 
pulse wavefront which causes the laser intensity to vary across the condensate.
In this case, different condensate atoms experience different pulse areas
which causes the ``angle'' of the Bragg pulse to vary.  That is, not all 
atoms will experience either a $\pi$ or $\pi/2$ pulse.  Our model can also
be generalized to handle Bragg pulses of arbitrary angle.  These ``angle''
errors in the pulses are analogous to imperfections in the thickness of the
blades in a neutron interferometer.  Such errors can be addressed by a new
AI design that implements this QIS idea of the ``power of one qubit''~\cite
{PhysRevLett.81.5672}.  The implementation of this idea for Bragg 
interferometers will be the subject of a forthcoming article.

Atom interferometry with Bose--Einstein condensates hold the promise
for applications in ultra--sensitive navigation and precision metrology.
We noted earlier the idea that, because of the connection between quantum
algorithms and multi--particle interferometers, there is great potential
for using the advances in QIS to inspire advances in precision interferometer
designs.  The Bragg AI prototyping method presented here represents a new tool for the rapid assessment of new Bragg AI designs.  In the future we intend 
to apply our tool to design AIs that use QIS ideas for decoherence avoidance
(e.g., decoherence--free subspaces, \cite{Benenti_Casati_Strini}) and 
minimization (e.g., the power of one qubit,\cite{PhysRevLett.81.5672}).

\begin{acknowledgments}
The authors acknowledge stimulating discussions with E. Hagley.  This 
work was supported by the NSF under grant numbers PHY--1004975, PHY--0758111, the Physics Frontier Center grant PHY--0822671 and by NIST.  The authors
also acknowledge assistance from Hadayat Seddiqi.
\end{acknowledgments}

\bibliography{bragg_prototyping_pra_revised}{}

\end{document}